\documentclass[aps,prx,twocolumn,floatfix,longbibliography,superscriptaddress]{revtex4-2}

\usepackage[T2A,T1]{fontenc}
\usepackage{amssymb, amsmath, amsthm}
\usepackage{graphicx}
\usepackage[dvipsnames]{xcolor}
\usepackage{soul}
\usepackage[colorlinks=true,citecolor=myblue,linkcolor=myred]{hyperref}
\definecolor{myblue}{rgb}{0.2,0.2,0.8}
\definecolor{myzard}{cmyk}{0,0,0.05,0}
\definecolor{mywhite}{rgb}{1,1,1}
\definecolor{myred}{rgb}{1,0.,0.3}
\makeatletter
\@ifundefined{textcolor}{}
{%
 \definecolor{BLACK}{gray}{0}
 \definecolor{WHITE}{gray}{1}
 \definecolor{RED}{rgb}{1,0,0}
 \definecolor{GREEN}{rgb}{0,1,0}
 \definecolor{BLUE}{rgb}{0,0,1}
 \definecolor{CYAN}{cmyk}{1,0,0,0}
 \definecolor{MAGENTA}{cmyk}{0,1,0,0}
 \definecolor{YELLOW}{cmyk}{0,0,1,0}
}

\makeatother

\usepackage{babel}

\begin{document}
\title{Dispersive readout of molecular spin qudits}
\author{\'{A}lvaro G\'{o}mez-Le\'{o}n}
\affiliation{Instituto de F\'{i}sica Fundamental IFF-CSIC, Calle Serrano 113b, Madrid 28006, Spain}
\author{Fernando Luis}
\affiliation{Instituto de Nanociencia y Materiales de Arag\'{o}n (INMA), CSIC-Universidad de Zaragoza, Zaragoza 50009, Spain}
\author{David Zueco}
\affiliation {Instituto de Nanociencia y Materiales de Arag\'{o}n (INMA), CSIC-Universidad de Zaragoza, Zaragoza 50009, Spain}
\date{\today}
\begin{abstract}
We study the physics of a magnetic molecule described by a "giant" spin with multiple ($d > 2$) spin states interacting with the quantized cavity field produced by a superconducting resonator. By means of the input-output formalism, we derive an expression for the output modes in the dispersive regime of operation. It includes the effect of magnetic anisotropy, which makes different spin transitions addressable. We find that the measurement of the cavity transmission allows to uniquely determine the spin state of the qudits. We discuss, from an effective Hamiltonian perspective, the conditions under which the qudit read-out is a non-demolition measurement and consider possible experimental protocols to perform it. Finally, we illustrate our results with simulations performed for realistic models of existing magnetic molecules.
\end{abstract}
\maketitle

\section{Introduction}

Circuit quantum electrodynamics (QED) studies the coupling of quantized cavity modes in superconducting resonators to "artificial atoms"~\cite{Wallraff2004,Blais2020}. It provides a practical method to readout the state of circuits involving Josephson junctions~\cite{Wallraff2005}. The readout protocol is based on measuring the shift of the cavity resonance frequency induced by its coupling to the qubit. When the two systems are energetically detuned from each other, a condition often referred to as the dispersive regime, the shift depends on whether the latter is in state '0' or '1'~\cite{Blais2004}. This technology has been applied in most of the currently available quantum processors that use superconducting qubits~\cite{Martinis2019,Cross2019}. 

In the last decade, circuit QED has been enriched with studies of hybrid platforms, in which electron~\cite{Ruggenthaler2018} and, particularly, spin ensembles~\cite{Xiang2013,Clerk2020} are coupled to on-chip superconducting cavities. Different magnetic species have been already studied in this context, including impurity spins in semiconductors~\cite{Kubo2010, Schuster2010,Amsuss2011,Weichselbaumer2019}, lanthanide ions~\cite{Bushev2011, Probst2014} and magnetic molecules~\cite{Bonizzoni2017,Mergenthaler2017,Gimeno2020,Bonizzoni2020}. A potential application of these schemes is the implementation of quantum memories~\cite{Blencowe2010}, exploiting the fact that some spin systems exhibit very long coherence times~\cite{Morello2014}. Yet, spins can also perform as operational qubits~\cite{Awschalom2013}. Hybrid processors based on microscopic spins coupled to, and through, on-chip resonators could outperform superconducting processors on account of their larger potential for integrating many resources in a single device~\cite{Jenkins2016,Morello2017,carretta2021}. 

\begin{figure}
\includegraphics[width=0.9\columnwidth]{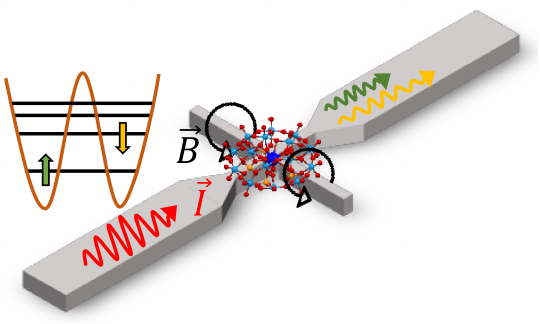}
\caption{Schematic setup where a magnetic molecule is placed on a constriction fabricated in the central line of a superconducting coplanar resonator. Sending microwave pulses and detecting transmission allows to differentiate between the multiple molecular spin states due to the different frequency shifts they generate on the cavity frequency.}
\label{fig:Schematic}
\end{figure}

In connection with the latter idea, a further appealing property of solid state spins is that they provide natural realisations of
qudits, i.e. quantum systems with $d>2$ discrete levels. The ability to use additional states provides resources for quantum
information processing~\cite{Lanyon2008,Campbell2014}, which can be especially useful for the implementation of specific quantum
error correction codes~\cite{Gottesman2001,Pirandola2008,Chiesa2020} or the quantum simulation of problems involving multiple
degrees of freedom~\cite{Tacchino2021}. Of particular relevance in this context are artificial magnetic molecules~\cite{Gaita2019,Sessoli2019} (see Fig.~\ref{fig:Schematic}), whose composition and structure can be chemically tuned to create
well-defined systems with multiple and experimentally accessible low-lying spin states~\cite{Luis2011,Aguila2014,Ferrando2016a,Jenkins2017,Moreno-Pineda2017,Moreno-Pineda2018,Luis2020,Gimeno2021}.
The challenge is finding practical methods to operate these multidimensional quantum spin systems.

In this work, we study the possibility of reading out the states of molecular spin qudits via their coupling to cavity photons
(see Fig.~\ref{fig:Schematic}). Our main goal is to generalise the dispersive readout protocol known for qubits to the case of
qudits. However, this generalisation is not straightforward. Magnetic molecules require some careful considerations, on account of
their inherent level multiplicity, the presence of a complex ligand field and, in the case of molecules hosting more than a single
magnetic ion, spin-spin interactions. The last two contributions make the different levels unequally spaced, thus allowing to
spectroscopically address each resonant transition. Yet, they also introduce some marked quantitative and qualitative differences
with respect to the case of qubits. These aspects determine how the readout process modifies the qudit state and open diverse
alternatives for its experimental implementation. Last but not least, since first realizations 
will likely involve experiments on molecular crystals, in order to enhance the collective spin-photon 
couplings, one needs to consider also the inhomogeneous broadening of the spin qudit levels.

In order to rigorously address these questions, we calculate the transmission of a cavity coupled to one or several spin qudits using the input-output formalism~\cite{Burkard2016,Kohler2018,Mi2018}. 
Then, we obtain analytical expressions for the (state-dependent) cavity frequency shift in the dispersive regime. Furthermore, using a Schrieffer-Wolff transformation, we derive an effective Hamiltonian for the dressed spin states, which describes the spin-photon coupled system. This allows to address the non-demolition nature of the spin state readout~\cite{Pereira2021}. The general theory is tested in four examples. First, we study a simple $S=1$ spin model with purely uniaxial anisotropy, which closely resembles the physics of N-V centers. Then, we move on to molecules that illustrate three different 
prototypical situations: a $\rm {GdW_{30}}$ qudit~\cite{Jenkins2017} based on a large $S=7/2$ electronic spin, the {[}CeEr{]} molecular dimer~\cite{Aguila2014}, which hosts two weakly interacting effective spin-1/2 systems, and the $^{173}\text{Yb-trensal}$, which couples an effective $S=1/2$ electronic spin to an $I=5/2$ nuclear spin.  

The remainder of the manuscript is organised as follows. Section~\ref{sec:Theory} is devoted to the general theory for  the coupling of molecular magnets and superconducting cavities or $LC$-resonators,  the dispersive readout formulas and the non-demolition character of the measurement.  Then, Section~\ref{sec:Results} reports the different examples where we apply the theory discussing the feasibility of our proposal.  We close the paper with some conclusions.  Technical details and concepts of minor importance are left for the appendices.


\section{Theory\label{sec:Theory}}
\subsection{Magnetic c-QED}

Most molecular magnets are both neutral and exhibit a close to zero electric dipole. In many cases, they are 
accurately described by a simple ``giant'' spin-$S$ effective Hamiltonian $\mathcal{H}_{S}$, which
includes the effects of magnetic anisotropy and the couplings to magnetic fields~\cite{gatteschi2006, bartolome2016}. This Hamiltonian is naturally written in terms of Stevens operators:
\begin{equation}
\mathcal{H}_{S}=\sum_{k=2,4,6}\sum_{q=-k}^{k}B_{k}^{q}\hat{O}_{k}^{q}\left(\vec{S}\right)+\mu_{B}\vec{B}\cdot\hat{g}\cdot\vec{S},\label{eq:Initial-H}
\end{equation}
where $B_{k}^{q}$ are magnetic anisotropy constants,  $\vec{S}=(S^{x},S^{y},S^{z})$ are spin operators with $[S^j, S^k] = i  \epsilon_{jkl} S^l$, $\mu_{\rm B}$ is the Bohr magneton, $\vec{B}=(b_x,b_y,b_z)$ the external magnetic field and $\hat{g}$ the gyromagnetic tensor. Due to the anisotropy terms in Eq.~\eqref{eq:Initial-H}, the molecule is characterised by a set of $2S+1$ unequally spaced energy levels which can be independently addressed and used to define a qudit~\cite{Muthukrishnan2000,Brennen2005,Jenkins2017,Luis2020,Kraus2007,Gimeno2021,carretta2021}. 
For simplicity, we will use Eq.~\eqref{eq:Initial-H} to define our qudit and to illustrate the theory describing the coupling to cavity photons. The same formalism can be adapted to more complex situations that fall beyond the giant spin approximation. The case of two weakly coupled spins is considered below for the molecular lanthanide dimer [CeEr], as well as the case of $^{173}\text{Yb-trensal}$, which contains electronuclear spin states involving the electronic and nuclear spins.

The coupling of molecular magnets to a superconducting cavity or LC-resonator is described by the generalised Dicke model, which reduces to the quantum Rabi model for the case of a single molecule~\cite{Jenkins2014,Jenkins2016}. 
This description can be further simplified to a Jaynes-Cummings model, by neglecting counter-rotating terms. However, as we are interested in the dispersive regime, which avoids resonances, and the role of counter-rotating terms can be relevant to some extent, we will keep them and use the more general description of the system.
The electromagnetic field in the cavity/LC-resonator is quantized and its Hamiltonian is given by:
\begin{equation}
\mathcal{H}_{c}=\Omega a^{\dagger}a \ ,
\end{equation}
with $\Omega$ its frequency and $a$($a^{\dagger}$) the photonic annihilation(creation) operators which fulfil $\left[a,a^{\dagger}\right]=1$. The local quantized magnetic field  generated by the superconducting currents  can be written as $\vec{B}_{\textrm{mw}}\left(\vec{r}\right)=\vec{B}_{\textrm{rms}}\left(\vec{r}\right)\left(a^{\dagger}+a\right)$, with $\vec{B}_{\textrm{rms}}^{2}\left(\vec{r}\right)=\langle0|\vec{B}_{\textrm{mw}}^{2}\left(\vec{r}\right)|0\rangle$ its zero-point fluctuations.
This quantum field couples to the spin via the Zeeman interaction and therefore adds a contribution analogous to the second term in Eq.~\eqref{eq:Initial-H}, with the difference that in this case there will be back-reaction between the spin and photon fields. Therefore, the total Hamiltonian is:
\begin{equation}
\mathcal{H}=\mathcal{H}_{S}+\mathcal{H}_{I}+\mathcal{H}_{c},\label{Hamiltonian-coupled}
\end{equation}
with
\begin{equation}
\mathcal{H}_{I}=\left(a^{\dagger}+a\right)\vec{\lambda}\cdot\hat{g}\cdot\vec{S},\label{eq:H_Interaction}
\end{equation}
and $\vec{\lambda}=\mu_{B}\vec{B}_{\textrm{rms}}\left(\vec{r}\right)$ the coupling constant for a molecule at position $\mathbf{r}$. 
The specific form of $\mathcal{H}_{I}$ will slightly change for the cases of the $\left[\text{CeEr}\right]$ dimer and the $^{173}\text{Yb-trensal}$ electronuclear molecule. However, its treatment is analogous in all cases.
\subsection{Dispersive readout and state dependent cavity transmission}

Dispersive readout was developed for the non-demolition readout of qubits. The main idea is to obtain information of a quantum system indirectly, by measuring its effect on the propagation of the cavity photons. In the dispersive regime, where the qubit is detuned from the cavity, the state of the former shows a one-to-one correspondence with the frequency shift of the latter, which can be determined by measuring the transmission through the device~\cite{Blais2004,Zueco2009}. Importantly, this readout is a non-demolition measurement~\cite{QND1,QND2}, which greatly improves the potential use of this technique. 
The generalisation of the readout protocol to qudits in magnetic molecules is not straightforward. The two main difficulties lie in the multilevel energy spectrum for spin $S>1/2$~\cite{Asenjo-Garcia2019} and in the presence of magnetic non-linear contributions from the anisotropy terms. We show now how to perform this generalization.

For simplicity, we first focus on the case of a single molecule interacting with the cavity field, but its generalisation to an ensemble of molecules is also addressed below. For our purposes, it is convenient to write the spin operators in Eq.~\eqref{Hamiltonian-coupled} in the basis of eigenstates of $\mathcal{H}_{S}$. In this basis, the Hamiltonian for the magnetic molecule takes a very simple form:
\begin{equation}
    \mathcal{H}_S = \sum_{\alpha=1}^{2S+1} E_{\alpha} X^{\alpha,\alpha}
\end{equation}
where $X^{\alpha,\alpha}=|\alpha\rangle\langle\alpha|$ is the projector onto the eigenstate $|\alpha\rangle$, with energy $E_{\alpha}$. Similarly, the interaction Hamiltonian, $\mathcal{H}_{I}$ in Eq.~\eqref{eq:H_Interaction}, takes the following form:
\begin{equation}
\mathcal{H}_{I}=\left(a^{\dagger}+a\right)\sum_{\vec{\alpha}=1}^{2S+1}\Lambda_{\vec{\alpha}}X^{\vec{\alpha}} ,
\label{eq:H_Interaction-1}
\end{equation}
where $X^{\vec{\alpha}}=|\alpha_{1}\rangle\langle\alpha_{2}|$ are Hubbard operators~\cite{HubbardOp}. 
The matrix $\Lambda_{\vec{\alpha}}$ can be explicitly written in terms of the $S^{z}$ eigenstates $|S,M\rangle$. However, to write its explicit form below, we will consider that the gyromagnetic tensor $\hat{g}$ is diagonal, with non-zero elements $\left(g_{x},g_{y},g_{z}\right)$. This assumption is not strictly necessary, but it is always possible for single ion magnets and highly reduces the number of terms. For molecules with several ions within the same cluster special care is required, but we will discuss below how to deal with them using the example of the molecular complex {[}CeEr{]}. 
After some algebra, $\Lambda_{\vec{\alpha}}$ reads:
\begin{eqnarray}
\Lambda_{\vec{\alpha}} & = & \lambda_{z}g_{z}\sum_{M=-S}^{S}Mc_{\alpha_{1},M}c_{\alpha_{2},M}^{\ast}\\
 &  & +\sum_{M=-S}^{S}\gamma_{S,M}\frac{\lambda_{x}g_{x}-i\lambda_{y}g_{y}}{2}c_{\alpha_{1},M+1}c_{\alpha_{2},M}^{\ast}\nonumber \\
 &  & +\sum_{M=-S}^{S}\gamma_{S,M}\frac{\lambda_{x}g_{x}+i\lambda_{y}g_{y}}{2}c_{\alpha_{1},M}c_{\alpha_{2},M+1}^{\ast}\nonumber 
\end{eqnarray}
with $\gamma_{S,M}=\sqrt{S\left(S+1\right)-M\left(M+1\right)}$ and
$c_{\alpha,M}=\langle\alpha|M,S\rangle$. 
The advantage of working with Hubbard operators is that the non-linear contributions from the anisotropy terms are nicely encoded in the eigenstates $|\alpha\rangle$. Now we discuss how to link the spectral features of the molecule with the cavity photons.

To explore the cavity transmission in a two-port setup (see schematic in Fig.~\ref{fig:Schematic}) we consider the input-output formalism~\cite{Gardiner1985}. Following the general theory, the cavity field is computed using a quantum Langevin-like equation where the input field is explicitly considered as a source:

\begin{equation}
\partial_{t} a=-i\left(\Omega-i\frac{\gamma}{2}\right)a-i\sum_{\vec{\alpha}=1}^{2S+1}\Lambda_{\vec{\alpha}}X^{\vec{\alpha}}-\sum_{l=1,2}\sqrt{\gamma_{l}}b_{\textrm{in},l}\label{eq:EOM1}
\end{equation}
Eq.~\eqref{eq:EOM1} describes the time evolution of a photon operator in a cavity with frequency, $\Omega$, and total cavity loss, $\gamma=\gamma_{1}+\gamma_{2}$ ($\gamma=\Omega/Q$, with $Q$ the quality factor). 
Here, $\gamma$ is expected to be small, in order to describe the Markovian environment produced by the resonator loses, and for the system to work in the dispersive regime. 
In addition, $b_{\textrm{in},l}$ is the input signal sent to the cavity/LC resonator through a transmission line
at port $l$.

To solve the equation of motion for the photon operator one needs to apply a truncation scheme, to find a solution for $X^{\vec{\alpha}}$. Otherwise, the equation of motion for $a$, couples to the equation of motion for $X^{\vec{\alpha}}$, which couples to many-body operators of the form $aX^{\vec{\alpha}}$ and $a^\dagger X^{\vec{\alpha}}$, and so on.
A natural choice for experiments is to consider that the interaction between a single spin and the cavity photons is small. In that case, when both sub-systems are coupled, they remain almost unaltered and the corrections can be expressed in power series of $\Lambda_{\vec{\alpha}}$.
Therefore, in this case the Heisenberg equation of motion for a generic Hubbard operator can be simplified to:
\begin{equation}
\dot{X}^{\vec{\alpha}}\simeq iE_{\vec{\alpha}}X^{\vec{\alpha}}+i\left(a+a^{\dagger}\right)\Lambda_{\alpha_{2},\alpha_{1}}\left(\langle X^{\alpha_{2},\alpha_{2}}\rangle-\langle X^{\alpha_{1},\alpha_{1}}\rangle\right)\label{eq:EOM2}
\end{equation}
being $E_{\vec{\alpha}}=E_{\alpha_{1}}-E_{\alpha_{2}}$ the energy difference. To obtain Eq.~\eqref{eq:EOM2} we have neglected terms of order two or higher in $\Lambda_{\vec{\alpha}}$. This makes the off-diagonal averages $\langle a^\dagger \rangle$, $\langle a \rangle$ and $\langle X^{\alpha,\beta} \rangle$ (for $\alpha\neq\beta$) to approximately vanish. 
Now Eq.~\eqref{eq:EOM1} and Eq.~\eqref{eq:EOM2} form a closed set which can be easily solved with a Fourier transform to frequency domain.
Importantly, notice that Eq.~\eqref{eq:EOM2} is valid for arbitrary dissipation, which means that the solution can be used to explore both, the weak and strong coupling regimes~\cite{Forn-Diaz2019,Perez-Gonzalez2021}.

Finally, in order to obtain more compact expressions, we consider that the molecules can be described by a density matrix which is diagonal in the basis of eigenstates $\rho=\sum_{\alpha=1}^{2S+1}p_{\alpha}X^{\alpha,\alpha}$. This is compatible with the weak coupling assumption and realistic in experiments due to the initial state preparation.
Finally, as the transmission is defined as $t_{c}=\langle b_{\textrm{out},2}\rangle / \langle b_{\textrm{in},1} \rangle$, we find from the input-output relations:
\begin{equation}
    t_{c}(\omega) = \frac{i\sqrt{\gamma_{1}\gamma_{2}}}{\Omega-\omega-i\frac{\gamma}{2}+\sum_{\vec{\alpha}}\frac{p_{\vec{\alpha}}\left|\Lambda_{\vec{\alpha}}\right|^{2}}{\omega+E_{\vec{\alpha}}+i\eta}}\label{eq:Transmission}
\end{equation}
where we have assumed that the input field enters via port $1$. Here, $\eta$ is the phenomenological broadening of the molecular spin levels and we have defined $\left|\Lambda_{\vec{\alpha}}\right|^{2} \equiv \Lambda_{\alpha_{1},\alpha_{2}}\Lambda_{\alpha_{2},\alpha_{1}}$ and $p_{\vec{\alpha}} \equiv p_{\alpha_{1}}-p_{\alpha_{2}}$. Eq.~\eqref{eq:Transmission} shows that the transmission depends on the state of the molecular spin and generalises the dispersive readout of qubits to qudits with non-linear terms. Furthermore, we can extract from Eq.~\eqref{eq:Transmission} the photon frequency shift measured at $\omega\sim\Omega$, for a molecule in state $\beta$ and with small spectral broadening $\eta$, which is given by:
\begin{equation}
\delta\tilde{\Omega}_{\beta}=2\sum_{\alpha=1}^{2S+1}\frac{\left|\Lambda_{\alpha,\beta}\right|^{2}E_{\alpha,\beta}}{\Omega^{2}-E_{\alpha,\beta}^{2}}.\label{eq:Frequency-shift}
\end{equation}

The theory presented so far applies to single molecules. However, in experimental setups it is common to consider crystals, thus ensembles of molecules. 
This is useful because the effective coupling and the signal it generates, is enhanced with the number of molecules $N$ that couple to the quantized field. 
In this case, Eq.~\eqref{eq:Transmission} is still valid, but the frequency shift acquires an enhancement proportional to $N$:
\begin{equation}
\label{eq:Frequency-shift-ensemble}
    \delta\tilde{\Omega}_{\beta}\left(\omega\right)=\sum_{i=1}^{N}\sum_{\alpha=1}^{2S+1}\frac{2\left|\Lambda_{i}^{\alpha,\beta}\right|^{2}E_{\alpha,\beta}}{\left(\omega+i\eta\right)^{2}-E_{\alpha,\beta}^{2}},
\end{equation}
with $\Lambda_{i}^{\alpha,\beta}$ the matrix elements for each molecule.
The detailed derivation for the case of inhomogeneous couplings and the discussion about the regime of validity are in the Appendix~\ref{app:ensemble}.

\subsection{Quantum non-demolition character}
\label{sec:QND}
As previously stated, one of the advantages of the dispersive readout is the Quantum Non-Demolition (QND) nature of the measurement. This property can be easily understood in the qubit case from the effective Hamiltonian, to second order in the
interaction, obtained using a Schrieffer-Wolff transformation~\cite{S-W-Transformation}. There, the qubit and the photon frequency
correction terms commute, indicating that a projective measurement of the photon frequency will not change the qubit state. This
result does not necessarily hold in the case of a qudit, specially due to the presence of magnetic anisotropy terms. We now derive
an effective Hamiltonian to second order in $\Lambda_{\vec{\alpha}}$, to check if the dispersive readout still is a QND
measurement for magnetic molecules.

The Schrieffer-Wolff transformation is defined in terms of a matrix $\mathcal{S}$~\cite{S-W-Transformation}:
\begin{equation}
\tilde{\mathcal{H}}=e^{\mathcal{S}}\mathcal{H}e^{-\mathcal{S}}=\mathcal{H}_{0}+\frac{1}{2}\left[\mathcal{S},\mathcal{H}_{I}\right]+\mathcal{O}\left(\Gamma^{3}\right),\label{eq:S-W1}
\end{equation}
where $\mathcal{H}_{0}=\mathcal{H}_{S}+\mathcal{H}_{c}$ and $\Gamma$ is a small parameter (see below) and we have imposed $\left[\mathcal{H}_{0},\mathcal{S}\right]=\mathcal{H}_{I}$ in Eq.~\eqref{eq:S-W1} to push the coupling term to second order. This last condition is met with the \emph{ansatz}:
\begin{equation}
\mathcal{S}=\sum_{\vec{\beta}=1}^{2S+1}\left(\Gamma_{+}^{\vec{\beta}}a^{\dagger}+\Gamma_{-}^{\vec{\beta}}a\right)X^{\vec{\beta}},\ 
\Gamma_{\pm}^{\vec{\beta}}=\frac{\Lambda_{\vec{\beta}}}{E_{\vec{\beta}}\pm\Omega}. \label{eq:S-W2}
\end{equation}
Notice that it has been possible to find this ansatz due to the use of Hubbard operators. Otherwise, the non-linear terms from the magnetic anisotropy would spoil its derivation.
After some algebra and the same approximations used to derive the cavity transmission $t_c(\omega)$, we find (detailed derivation in the Appendix~\ref{app:nondem}):
\begin{eqnarray}
\tilde{\mathcal{H}} & \simeq & \sum_{\alpha=1}^{2S+1}E_{\alpha}X^{\alpha,\alpha}+\Omega a^{\dagger}a\label{eq:Effective-H}\\
 &  & +\sum_{\alpha,\beta=1}^{2S+1}\frac{\left|\Lambda_{\alpha,\beta}\right|^{2}}{E_{\alpha,\beta}-\Omega}\left(1+a^{\dagger}a\frac{2E_{\alpha,\beta}}{E_{\alpha,\beta}+\Omega}\right)X^{\alpha,\alpha}.\nonumber 
\end{eqnarray}
Equation~\eqref{eq:Effective-H} is derived under the assumption of small interaction, with respect to the qudit levels detuning from the cavity (i.e., $\Lambda_{\vec{\beta}} < ||E_{\vec{\beta}}|-\Omega|$), and is valid to describe the weak and strong coupling regimes~\cite{Forn-Diaz2019}.
Also, Eq.~\eqref{eq:Effective-H} perfectly agrees with the cavity frequency shift predicted for the qudit state $\beta$ in Eq.~\eqref{eq:Frequency-shift}. 
Finally, the effective Hamiltonian also encodes some extra information, such as the effect of magnetic anisotropy in the effective spin-photon interaction and the energy shift of each individual qudit level.

To check the QND nature of the readout, we calculate the commutator between the molecule Hamiltonian $\mathcal{H}_S$ and the frequency shift term in the effective Hamiltonian, which we denote as $\tilde{\mathcal{V}}$. 
As the effective Hamiltonian in Eq.~\eqref{eq:Effective-H} is obtained by keeping diagonal terms only, both terms commute and we could conclude that the readout is non-demolition. 
This is true for sufficiently short time-scales, where diagonal contributions dominate. 
However, the off-diagonal terms, previously discarded to obtain Eq.~\eqref{eq:Effective-H}, can become relevant at longer times, if e.g. the measurement protocol is relatively slow. 
If these terms are kept, we find that in general:
\begin{equation}
\left[\mathcal{H}_{S},\tilde{\mathcal{V}}\right]=\sum_{\beta_{i}=1}^{2S+1}E_{\vec{\beta}}\Phi_{\vec{\beta}}X^{\vec{\beta}},\label{eq:QND}
\end{equation}
where $\Phi_{\vec{\beta}}$ is a scalar function of the state $\beta$ (see Appendix~\ref{app:nondem} for details). This breakdown of the QND measurement can happen even for a qubit, if the interaction with the photon field is not completely orthogonal to the qubit quantization axis. However, the qudit case introduces some additional ingredients. As we show below for the specific case of a molecule with $S=1$, even for a perfectly orthogonal photon field, the crystal anisotropy introduces a non-vanishing correction. Nevertheless, even in this case it is possible to cancel this effect by applying suitably aligned DC magnetic fields.
\section{Results and discussion}
\label{sec:Results}

Our results reproduce the well-known expressions for the qubit case ($S=1/2$)~\cite{Zueco2009} and the effective Hamiltonian is explicitly obtained in the Appendix~\ref{app:2ls}. 
In what follows, we discuss three relevant examples to explore the performance of our proposal for the state-dependent readout of molecular qudits.
\subsection{Toy model~($S=1$)}
Probably the simplest generalization to qubits ($S=1/2$) is the case of a qutrit ($S=1$) with uniaxial anisotropy, a magnetic field aligned with the anisotropy axis and a purely transverse coupling $\vec{\lambda}=\left(\lambda_x,0,0\right)$ [Cf.~Eqs.~\eqref{eq:Initial-H},~\eqref{Hamiltonian-coupled} and~\eqref{eq:H_Interaction-1}]:
\begin{equation}
\mathcal{H} = D \left( S^{z} \right)^{2} + \xi_{z}S^{z} + \Omega a^{\dagger}a + \lambda_x g_{x}\left(a^{\dagger}+a\right)S^{x}.
\label{eq:toy}
\end{equation}
Here, $D$ is the second order magnetic anisotropy constant and $\xi_{z} = g_{z} \mu_{\rm B} B_{z}$. A key advantage of this model is that the eigenvectors of ${\mathcal H}_S$ are spanned by the $S^z$-basis. The energies are $E_{\pm}=D \pm \xi_{z}$ and $E_{0}=0$, while  $\Lambda_{\vec{M}}=\lambda_x g_{x}/\sqrt{2}$ for $M_{1}-M_{2}=\pm 1$, and zero otherwise. As a consequence, in this case the photon frequency shifts can be analytically obtained:
\begin{align}
\delta\tilde{\Omega}_{\pm} & =\frac{\lambda_x^{2}g_{x}^{2}E_{\pm}}{E_{\pm}^{2}-\Omega^{2}},\\
\delta\tilde{\Omega}_{0} & =-\lambda_x^{2}g_{x}^{2}\left(\frac{E_{+}}{E_{+}^{2}-\Omega^{2}}+\frac{E_{-}
}{E_{-}^{2}-\Omega^{2}}\right) \; ,
\end{align}
and importantly, as the shifts are state dependent, in theory the molecular spin states could be resolved spectroscopically.

\begin{figure}
\includegraphics[width=1.0\columnwidth]{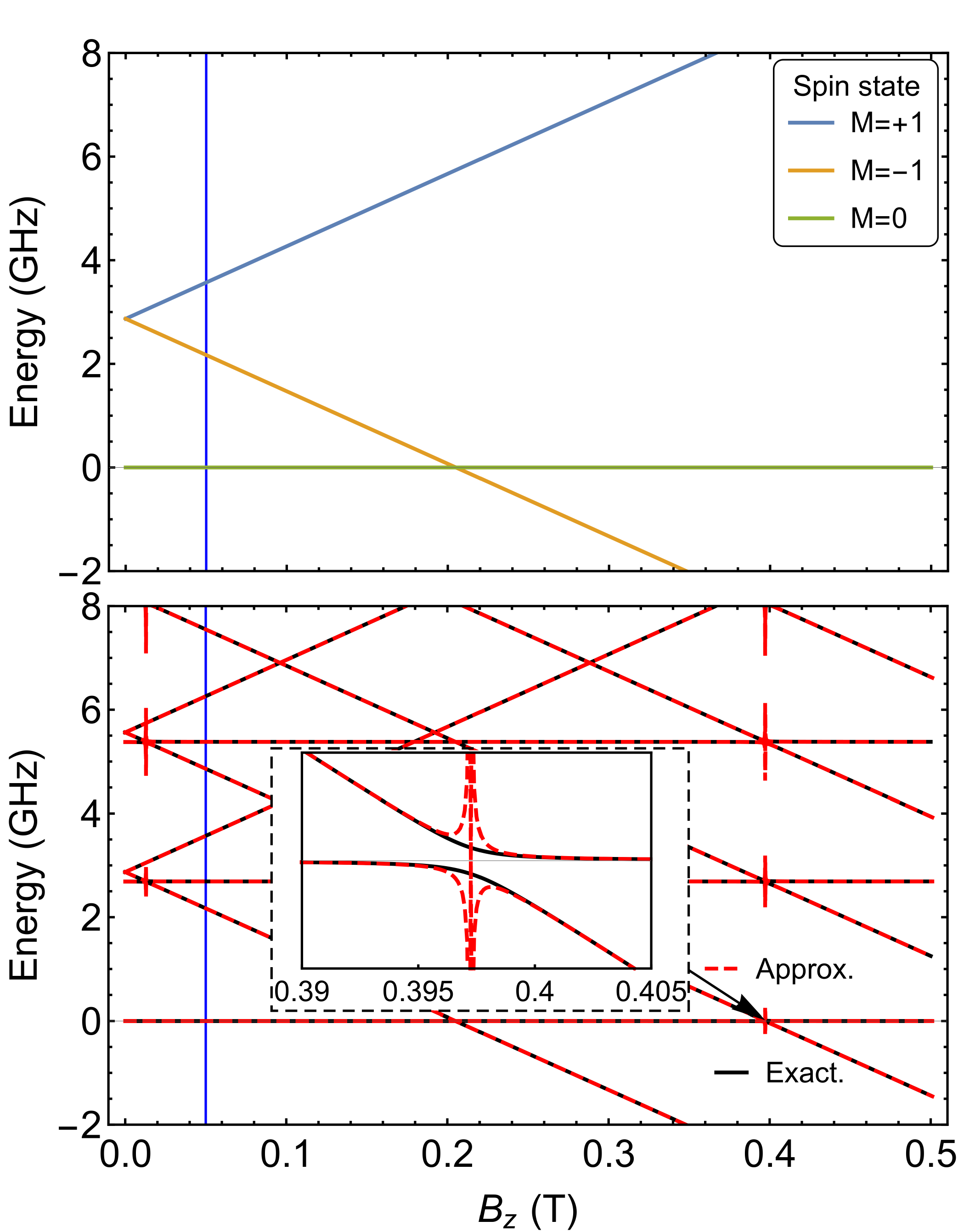}
\caption{\label{fig:Fig1}(Top) Energy levels for an isolated $S=1$ molecule (NV center) with a purely uniaxial anisotropy.
(Bottom) Energy levels for the spin-cavity system with resonator frequency $\Omega\simeq 2.69$GHz. The spectrum (solid black line) from exact diagonalization of Hamiltonian~\eqref{eq:toy} is compared to the approximation using Eq.~\eqref{eq:Effective-H}~(red dashed line). The inset zooms the region where the system is not in the dispersive regime due to a resonance between the qutrits and cavity.}
\end{figure}

NV centers~\cite{Doherty2012} in diamond provide an interesting physical realization of an $S=1$ system that can be accurately described by this simple model. The three lowest energy states can be described with a magnetic anisotropy  parameter $D\simeq 2.87$GHz. This is shown in Fig.~\ref{fig:Fig1}~(top), where we plot the energy levels of an isolated NV center, as a function of the longitudinal field $B_z$.

In this system, each single NV center weakly couples to photons. However, the strong-coupling regime has been achieved for the ensemble of NV centers in diamond crystals~\cite{Putz2014}, with collective coupling strengths in the MHz range. Here, we take experimental parameters reported in~\cite{Putz2014}, in order to check the feasibility of the readout in a realistic setup.
To find a range of parameters in which the spin-cavity system lies in the dispersive regime, in Fig.~\ref{fig:Fig1}~(bottom) we compare the exact spectrum of Eq.~\eqref{eq:toy} with the effective Hamiltonian from  Eq.~\eqref{eq:Effective-H}, for a collective coupling in the MHz range.
The plot shows very good agreement between the two, with the exception of regions where spins and cavity are resonant, as shown in the zoomed area near a resonant anti-crossing. The vertical blue line indicates our choice of the magnetic field to perform the readout.
%
\begin{figure}
\includegraphics[width=1.0\columnwidth]{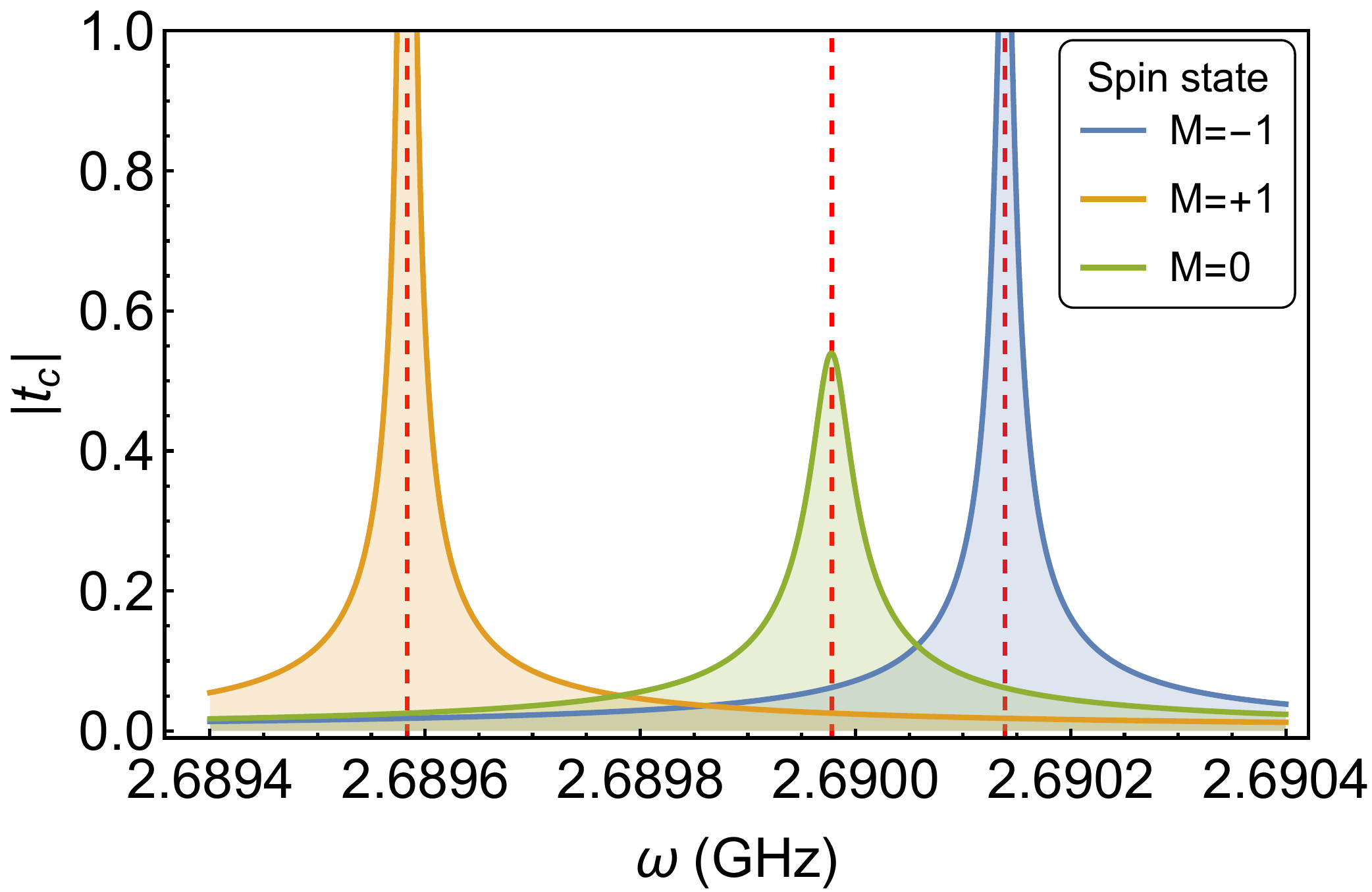}
\caption{\label{fig:Fig2} $\left|t_{c}\right|$ vs $\omega$ for different states of an ensemble of $S=1$ qudits or NV centers ($d=3$) and $\gamma_j=4\times10^{-5}$GHz. The inset shows the transmission phase $\phi$. Vertical dashed lines indicate the predicted frequency shift from Eq.~\eqref{eq:Frequency-shift}. The difference in amplitude between peaks is produced by the dissipative terms and the detuning of each transition from $\Omega$.}
\end{figure}

In experiments with NV centers \cite{Putz2014}, the resonator frequencies are close to $D$. Here, we take $\Omega= 2.6899$GHz and the inhomogeneous spin broadening $\eta\simeq 9.4$MHz. When the strong coupling regime is attained, the estimated value of the collective coupling results in $g_x\lambda_x\sqrt{N}\simeq 19.2$MHz. With these values, we can discuss the readout of the qutrit state in ensembles of NV centers.

Figure~\ref{fig:Fig2} shows the transmission amplitude as a function of frequency for the different spin states of a $S=1$ molecule or an NV center.
The static longitudinal field is marked with a vertical blue line in Fig.~\ref{fig:Fig1}, where the dispersive regime is well justified.
As a further numerical check, we have confirmed that the position of the peaks in Fig.~\ref{fig:Fig2} agrees with the peaks from the spectral function, calculated with exact diagonalization. 
The inset shows the transmission phase, which also depends on the spin state.
Phase acquisition has the advantage of having less fluctuations so it can be measured more accurately.
It also provides a quite direct method to determine the qudit spin state by measuring its sign on both sides of the central frequency and applying the truth table $++ \longrightarrow M = -1$, $+- \longrightarrow M = +1$ and $-- \longrightarrow M = 0$.

Importantly, for illustrative purposes, we have considered in Fig.~\ref{fig:Fig2} a resonator with losses $\gamma_j\simeq 10^{-5}$GHz. This value is smaller than the one considered in ref~\cite{Putz2014}, but it is still realistic. 
The reason is that for larger losses, the peaks in Fig.~\ref{fig:Fig2} broaden and overlap, however, this does not mean that the state cannot be detected, as the phase measurement remains almost unchanged. 
This means that the qudit readout is not highly constrained by the spin broadening (which is much larger in this case) and that phase measurements are robust, even for realistic resonators and in the presence of a sizeable inhomogeneous broadening.

This example demonstrates that the dispersive readout of an ensemble of molecules or other spin systems with $S>1/2$ and magnetic anisotropy is feasible. 
However, an important difference is that the characterisation of the spin qudit state requires to perform $d-1$ measurements. 
Nevertheless, this could be experimentally mitigated by considering multi-frequency pulses such as frequency combs that have been applied to multiplex the readout of arrays of $LC$ resonators used as radiation detectors in Astronomy~\cite{Rantwijk2016}. The pulse design can be likely optimized via the application of optimal control techniques, just as it is done with the spin control pulses~\cite{Castro2021}.

As we discussed in Section~\ref{sec:QND}, another difference with respect to qudits is the QND character of the readout. In qudits it is QND to the extent that off-diagonal contributions in Eq.~\eqref{eq:Effective-H} can be neglected. Yet, these terms slightly rotate the eigenstates of the isolated spin Hamiltonian with respect to those of the frequency shift term. If off-diagonal terms are included, the relevant commutator  in Eq.~\eqref{eq:QND} yields:
\begin{equation}
\label{eq:qndtoy}
\sum_{\beta_{i}=1}^{2S+1}E_{\vec{\beta}}\Phi_{\vec{\beta}}X^{\vec{\beta}} = \left(X^{+,-}-X^{-,+}\right) \sum_{\alpha=\pm}\frac{(\lambda_x g_x)^2 E_{\alpha}}{E_{\alpha}^{2}-\Omega^{2}}.
\end{equation}
Notice that the deviation from an ideal non-demolition measurement is, as we anticipated, a consequence of the magnetic anisotropy $D$. This can be seen if we take the limit $D\to0$ in Eq.~\eqref{eq:qndtoy}, which makes it vanish for $E_{\pm}=D\pm\xi_z$. 
Nevertheless, the correction to an ideal QND measurement is small, as it is proportional to $\lambda_x^2$ and to the average of  the off-diagonal operators, $X^{\pm,\mp}$, which involve two spin-flip processes (also proportional to $\lambda_x^2$). 
Therefore,  corrections become important only beyond the dispersive regime or for long time-scales (i.e., when the spins have evolved in time, moving significantly away from the initial state, $\rho$, that is to be measured), and for all practical purposes one can safely consider that the transmission readout is a QND measurement.
Interestingly, it would also be possible to suppress this effect in some cases by aligning the external magnetic field along specific directions, in order to make Eq.~\eqref{eq:QND} vanish. Concretely, in this case the working point $\xi_z=\sqrt{D^2 - \Omega^2 }$ makes Eq.~\eqref{eq:qndtoy} go to zero.

We have presented a simple model where the fundamental features of magnetic c-QED can be grasped in simple terms. However, models of artificial magnetic molecules can be more complex and contain a larger number of levels. For this reason we now consider three cases of fundamentally different artificial magnetic molecules, which are currently under study as interesting candidates for quantum technologies.\\

\subsection{Single ion magnet $\textrm{Gd}\textrm{W}_{30}$}

The inorganic molecular moiety $\textrm{Gd}\textrm{W}_{30}$ encapsulates a single $S=7/2$ Gd$^{3+}$ ion with long spin coherence time. Gadolinium has some characteristic traits that make this system of particular interest: it has the largest spin of the
periodic table and, because of its close to spherical $4f$ electronic shell, the zero field splittings between spin levels are
one to two orders of magnitude smaller than those found for other lanthanide or transition-metal ions. These properties combined
provide a large set of levels with energies lying within the reach
of microwave cavities. Besides, and due to its weak but yet non-zero
magnetic anisotropy, different spin transitions have also different resonant frequencies and they can be independently addressed.
Finally, any operation can be implemented by concatenation of the subset of transitions that can be induced by resonant microwave pulses. As a consequence, this molecule is equivalent to
a universal $3$-qubit processor (since $2^{3}=2\times7/2+1$)~\cite{Jenkins2017}. 
\begin{figure}
\includegraphics[width=1.0\columnwidth]{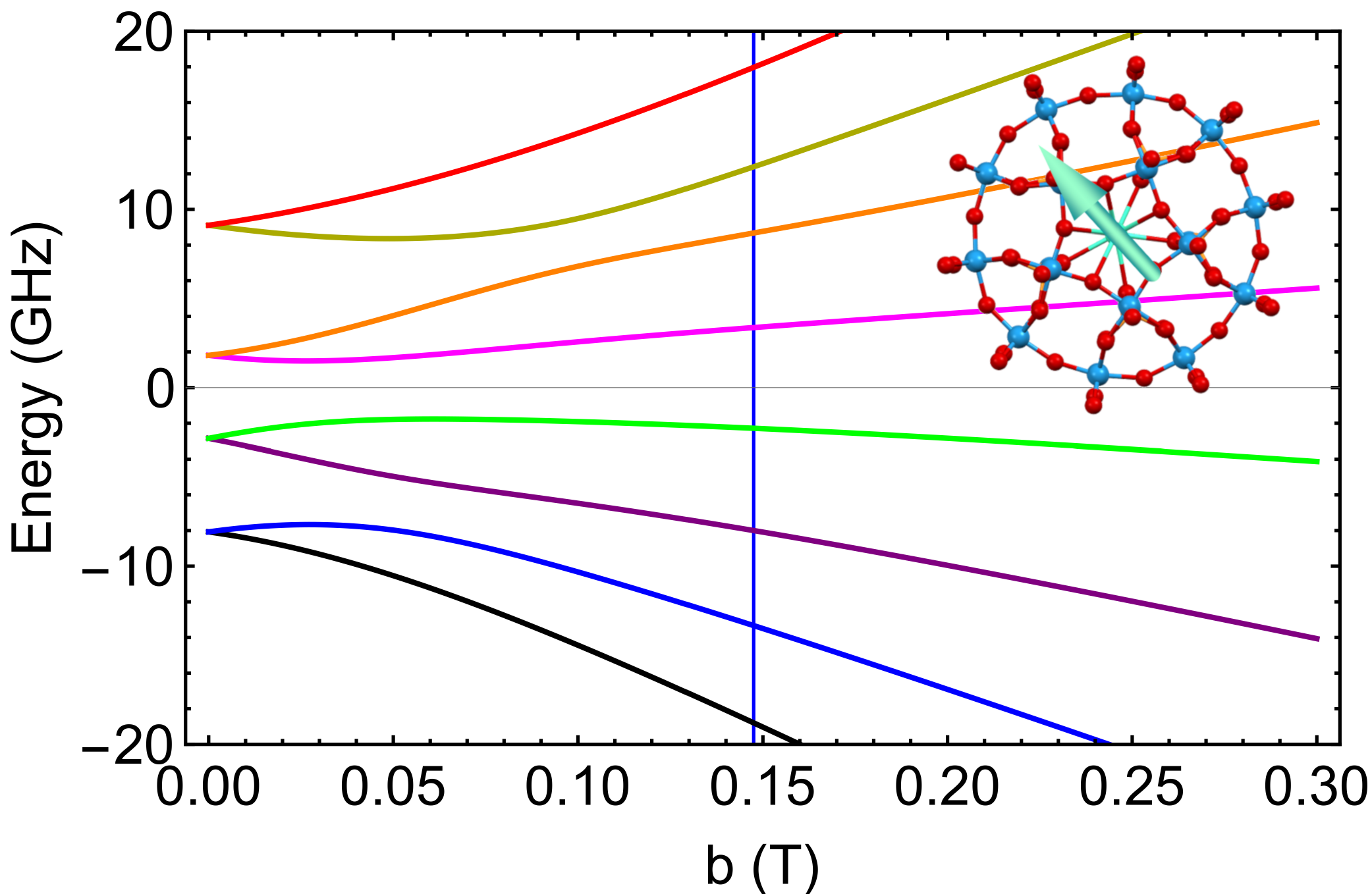}
\includegraphics[width=1.0\columnwidth]{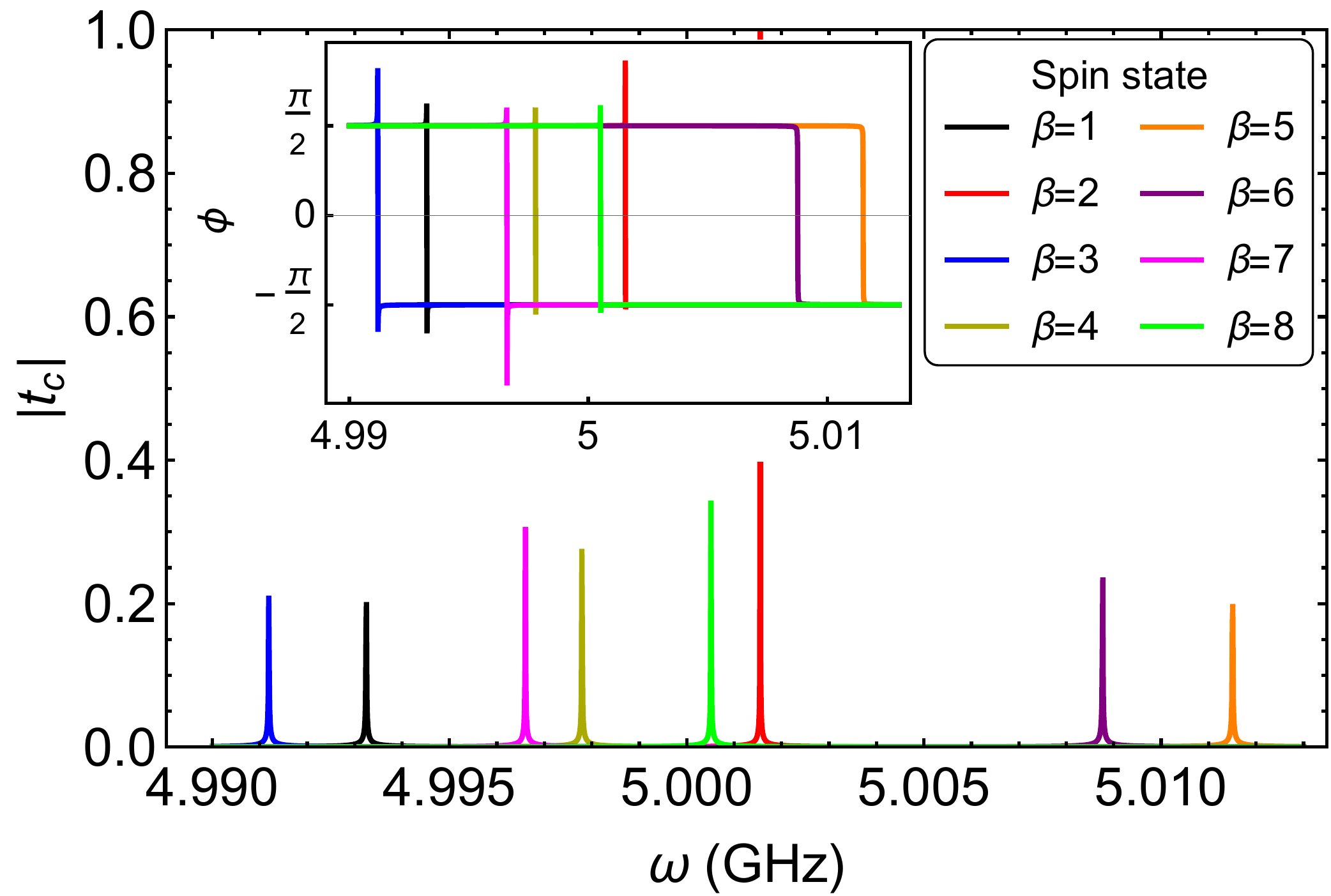}
\caption{\label{fig:Fig4} Top: Spin energy levels of $\text{GdW}_{30}$, shown in the inset, as a function of magnetic field $b$. The vertical blue line indicates the value chosen to calculate the transmission. Bottom: Transmission vs $\omega$ for the different spin states of a crystal of $N=1.6\times10^{14}$ $\text{GdW}_{30}$ molecules and a cavity frequency $\Omega=5$ GHz. The inset shows the phase of the transmission for the same range of $\omega$. The photon and spin decoherence rates take values $\gamma_{i}=10^{-6}$ GHz and $\eta=10^{-4}$ GHz.}
\end{figure}
Here, we show that, in addition to all this, every state can be resolved
by means of our dispersive readout proposal. The spin Hamiltonian of  $\textrm{Gd}\textrm{W}_{30}$ has longitudinal as well as
in-plane magnetic anisotropy terms. It can be written as:
\begin{equation}
\mathcal{H}_{S}=\frac{D}{3}O_{2}^{0}+E O_{2}^{2}-\mu_{B} g \vec{B}\cdot\vec{S},
\end{equation}
with $O_{2}^{0}=3\left(S^{z}\right)^{2}-S\left(S+1\right)$ and $O_{2}^{2}=\left(S^{x}\right)^{2}-\left(S^{y}\right)^{2}$.
Experimental values are $D=1.281\textrm{GHz}$, $E=0.294\textrm{GHz}$
and $g=2$. Typical spin decoherence times for this molecule are of the order of a few $\mu$s. 
This needs to be compared with the spin-photon
coupling. The magnetic field produced by the vacuum fluctuations in a standard design of a superconducting resonator is around $0.1$nT. This gives
an effective spin-photon coupling to
each of the allowed spin transitions of the order of a few Hz, which
is much smaller than the
value required to discriminate resonances associated with different spin states in the dispersive regime. 
However, this problem can be mitigated by considering ensembles of molecules, which provide the scaling $\sqrt{N}$ in the effective coupling [Cf. Eq.~\eqref{eq:Frequency-shift-ensemble}]. 
Considering a diluted single crystal containing only $1$\% of magnetic 
molecules dispersed in a diamagnetic host (for instance, the YW$_{30}$ derivative), the cavity would couple 
to approximately $1.6\times10^{14}$ molecules, providing collective  spin-photon couplings of the order of 
tens of MHz. 

Figure~\ref{fig:Fig4}~(top) shows the energy levels as a function of the magnetic field. In Fig.~\ref{fig:Fig4}~(bottom) the transmission is plotted as a function of $\omega$ for a cavity with frequency $\Omega=5$ GHz and a field configuration $\vec{B}=(1,0.3,0.3)b$  with $b=0.1475$T [vertical line in Fig.~\ref{fig:Fig4}~(top)]. It shows that the transmission resonance is different for each state of the molecule and that the frequency shifts can be experimentally resolved, provided that the spin levels are 
predominantly homogeneously broadened. This is crucial for molecules with a large number of levels, where frequency crowding can hinder resolving between states and lead to similar transmission resonances. It is also very important to determine an adequate field configuration $\vec{B}$, where $d-1$ frequency shifts are sufficiently large. Clearly, this highly depends on the magnetic anisotropy of the molecule and on the available frequencies of the resonator $\Omega$. Nevertheless, if two or more levels cannot be resolved, repeating the process at different magnetic field values would allow to completely determine the spin state. The inset in Fig.~\ref{fig:Fig4} shows again that the phase of the transmission can be used to perform the readout.
\begin{figure}
    \includegraphics[width=1\columnwidth]{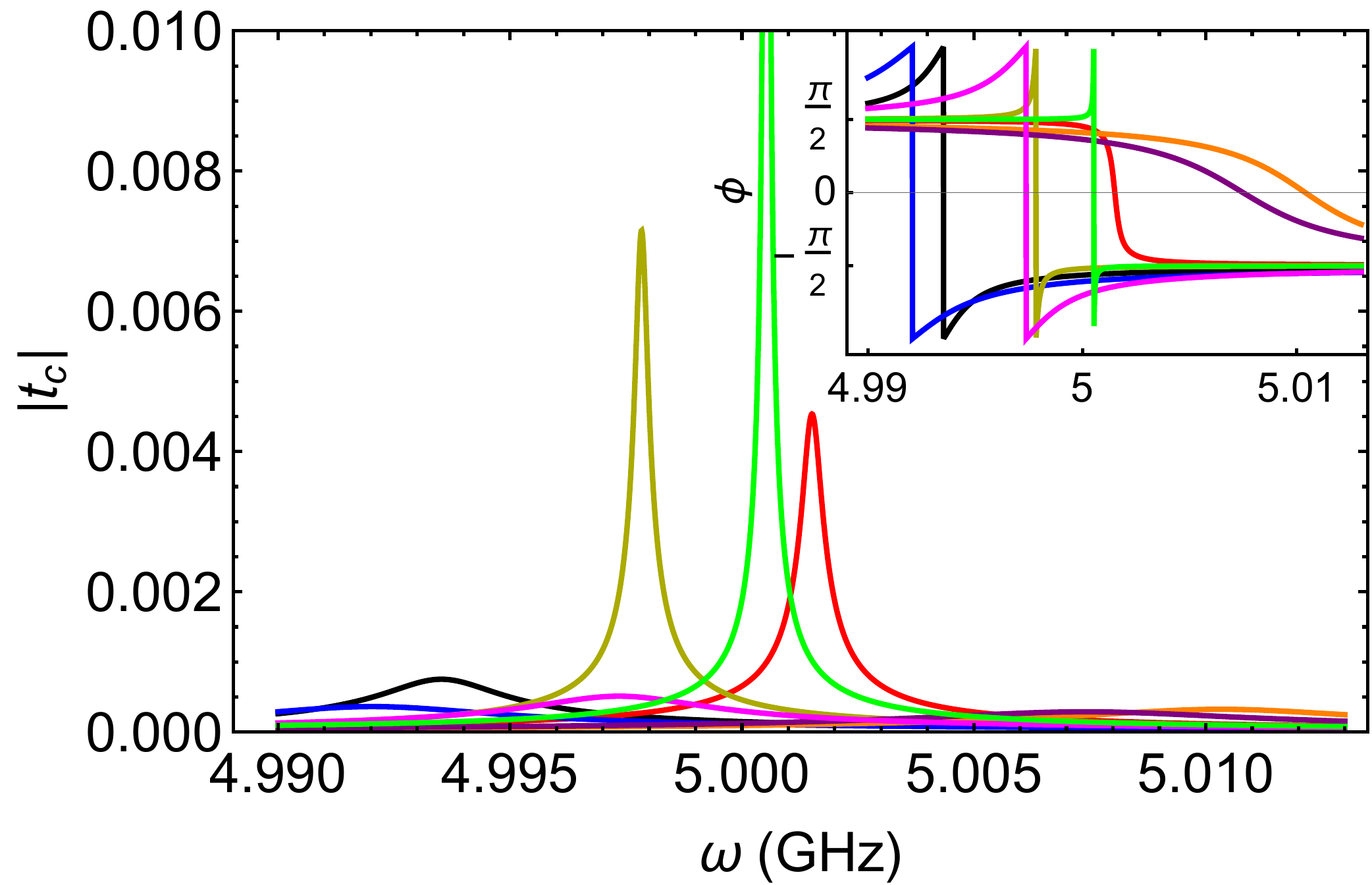}
    \caption{$|t_c|$ vs $\omega$ for the same parameters as in Fig.~\ref{fig:Fig4}, but with $\eta=0.1$GHz, which is the inhomogeneous broadening of the spin levels determined for diluted samples of $\text{GdW}_{30}$ \cite{Jenkins2017}. The transmission peaks are highly suppressed and they partially overlap. Still, information about the state can be extracted from the transmission phase (inset).}
    \label{fig:GdW-Broadening}
\end{figure}

Unfortunately, in this particular molecular system the inhomogeneous broadening remains large ($>100$ MHz~\cite{Jenkins2017}) even for highly diluted crystals, probably as a result of strains in the magnetic anisotropy parameters associated with the presence of different metastable molecular conformations at low temperatures. 
As can be seen in Fig.~\ref{fig:GdW-Broadening} (cf Fig.~\ref{fig:Fig4}), the main consequences are the broadening and amplitude reduction of the peaks, as it was previously seen in Fig.~\ref{fig:Fig2}.
This is due to the strength of dissipative terms and the detuning of each transition from the resonator frequency.
This limits, although it does not completely preclude, the ability to discriminate between different spin states. 

Nevertheless, it is important that these results show what the dominant sources of error are, to direct future research towards magnetic molecules with adequate properties for their manipulation in cavities.
\subsection{Heterodimetallic {[}CeEr{]} Lanthanide Complex}
The molecular dimer {[}CeEr{]} behaves, at sufficiently low $T$,
as two weakly coupled anisotropic $S=1/2$ spins, i.e., it deviates from the giant spin approximation. It provides a model situation to explore a system of two dissimilar, thus addressable, spin
qubits. Their mutual coupling allows to implement conditional two-qubit gates and ensures universal operations within the $d=4$ Hilbert space~\cite{Aguila2014}.
\begin{figure}
\includegraphics[width=1.0\columnwidth]{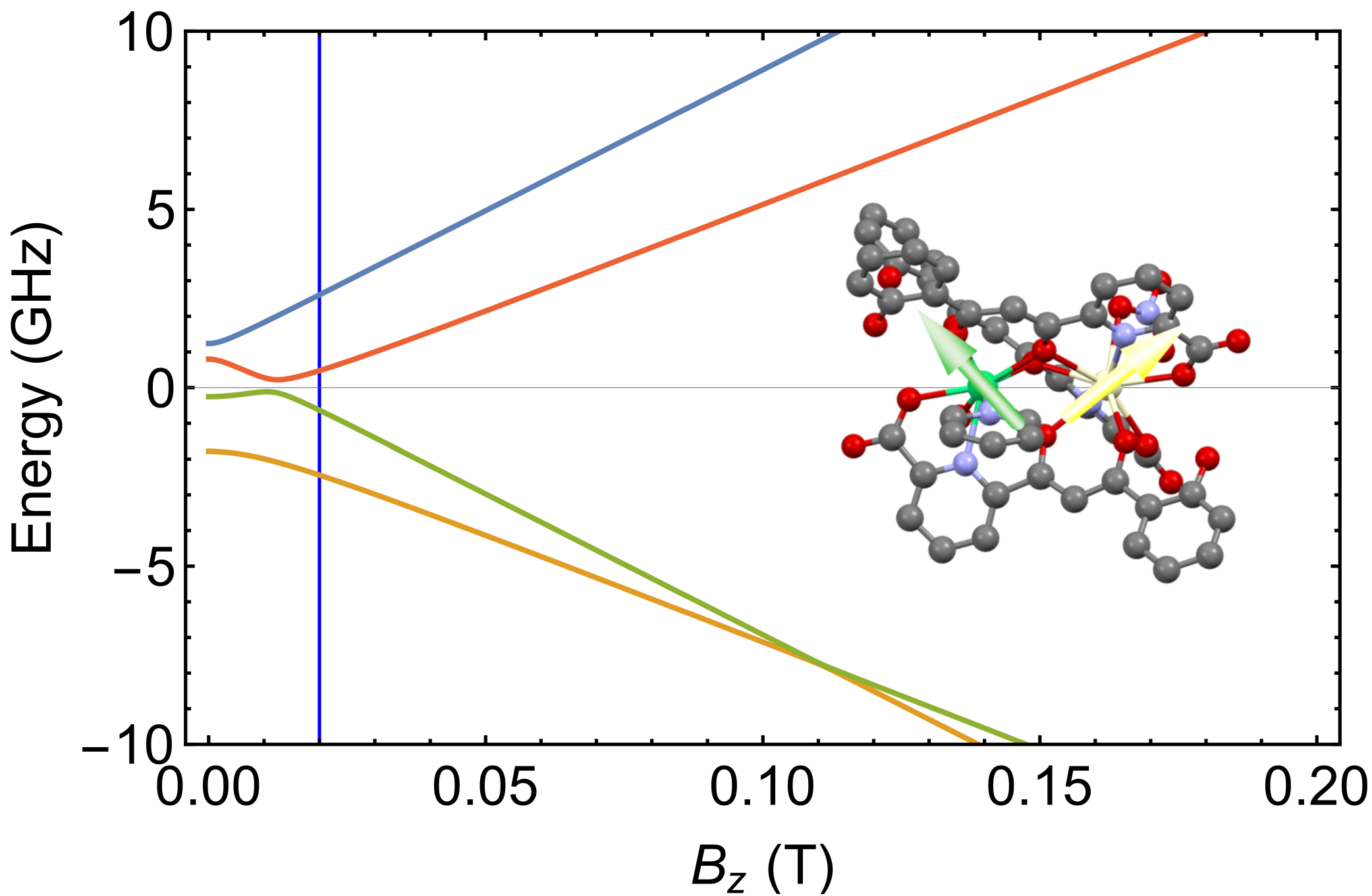}
\includegraphics[width=1.0\columnwidth]{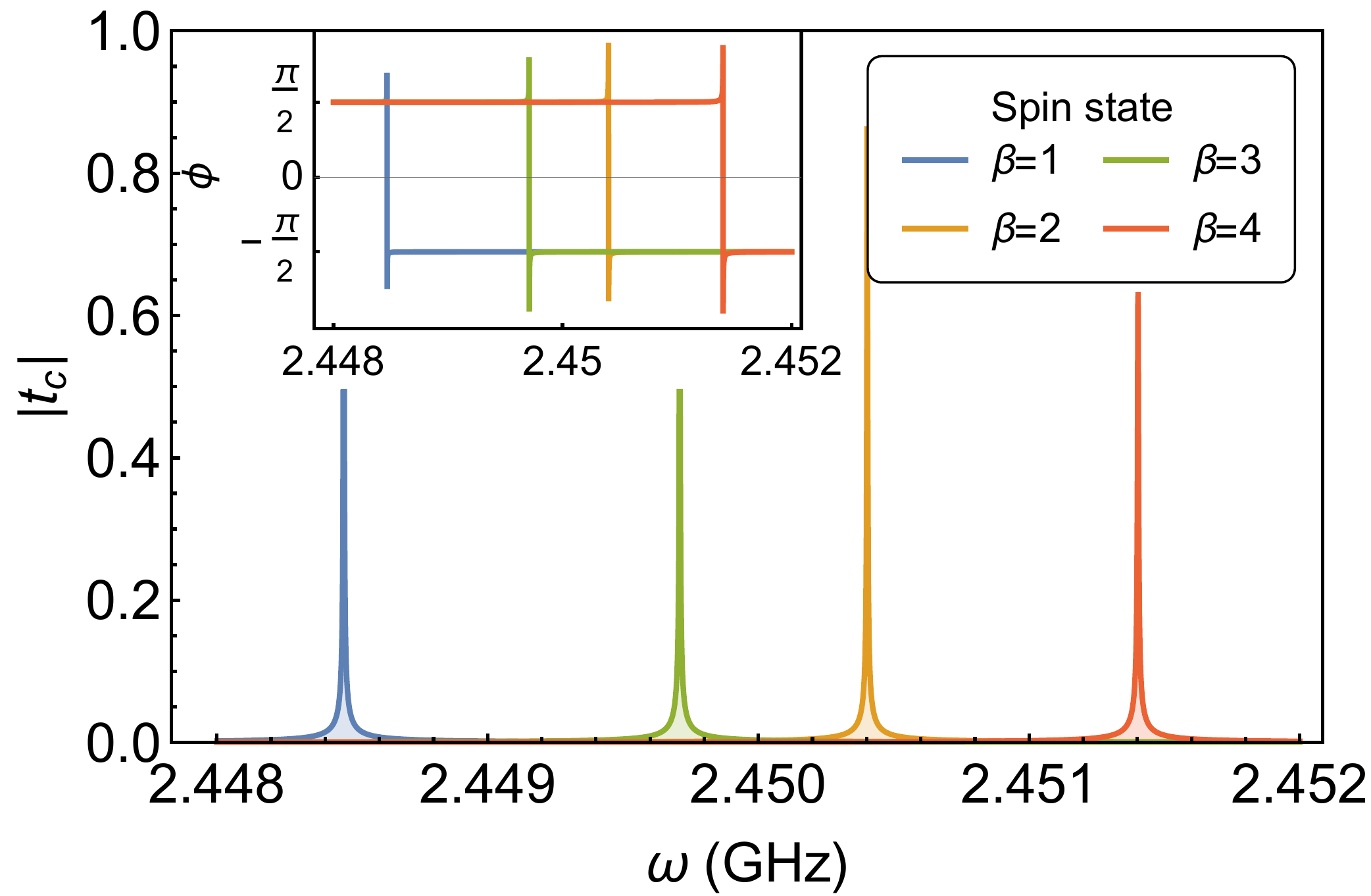}
\caption{\label{fig:Fig5} Top: Spin energy levels of {[}CeEr{]}, shown in the inset, as a function of magnetic field. Bottom: Transmission $\left|t_{c}\right|$ vs $\omega$ for a cavity frequency $\Omega=2.45$GHz and $B_z=0.02$T. The inset shows the phase of the transmission for the same frequency range. Parameters: $N=2.7\times 10^{14}$, $\gamma_{i}=10^{-6}$GHz and $\eta=10^{-4}$GHz.}
\end{figure}
This magnetic molecule can be described by the following spin Hamiltonian~\cite{Aguila2014}:
\begin{equation}
\mathcal{H}_S=-\mu_{B}\sum_{i=1,2}\vec{B}\cdot\hat{g}_{i}\cdot\vec{S}_{i}-\frac{J_{12}}{g_{J1}g_{J2}}\left(\hat{g}_{1}\cdot\vec{S}_{1}\right)\cdot\left(\hat{g}_{2}\cdot\vec{S}_{2}\right)
\end{equation}
where the effective gyromagnetic tensor of the $i$-th spin ($i=1,2$
for Er and Ce, respectively) is given by: $\hat{g}_{1}=\left(1.8,3.7,10\right)$
and $\hat{g}_{2}=\left(1,1.75,2.67\right)$, and the Landé factors
are $g_{J1}=6/5$ and $g_{J2}=6/7$, respectively. For the spin-spin interaction, we have chosen a scalar form of the interaction tensor with $J_{12}/k_{B}=-0.015\textrm{K}$,
and for the field configuration we choose $\vec{B}$ to be aligned
along the $z$-axis of the Er spin. The anisotropy axis of the Ce spin then makes an angle of about $\theta=70\text{º}$ with respect to $\vec{B}$. For simplicity, we choose
our frame of reference in such a way that the angle $\theta$ lies
in the x-z plane. The dimer levels are plotted in Fig.~\ref{fig:Fig5}~(top), as a function of the magnetic field.

When coupling to the cavity, the dimer molecule turns out to be slightly more complex than the previous discussed examples. This is due to the misalignment between the local anisotropies of both ions, which makes them couple differently with the cavity photons. The total
Hamiltonian can be written as follows (explicit expressions for each ion can be found in the Appendix~\ref{app:ceer}):
\begin{equation}
\mathcal{H} = \mathcal{H}_S+\mathcal{H}_c+\left(a^{\dagger}+a\right)\sum_{\vec{\alpha}}X^{\vec{\alpha}}\sum_{i}\vec{\epsilon}\cdot\hat{g}_{i}\cdot\langle\alpha_{1}|\vec{S}_{i}|\alpha_{2}\rangle
\end{equation}
where now $\hat{g}_{2}$ is a non-diagonal matrix and $X^{\vec{\alpha}}$
are the exact eigenstates of the isolated dimer (i.e., they are
many-body states of the two spins). Notice that in this case the effective
$\hat{g}$ tensor measured in an experiment would be very different
from the one of the isolated ions and that it contains contributions
from the interaction $J_{12}$. As in the previous case for the $\text{GdW}_{30}$ molecule, the coupling between a single molecule
and the cavity is too small, and we must consider ensembles of molecules.

In Fig.~\ref{fig:Fig5} we show the transmission as a function of 
$\omega$ for a cavity of frequency $\Omega=2.45$GHz and longitudinal field $h_z=0.02$T. It shows that the frequency shift is large enough to 
be experimentally resolved. The inset shows that the phase $\phi$, again, can be used for unequivocally determining the spin state of the molecule. 
\begin{figure}
    \centering
    \includegraphics[width=1.0\columnwidth]{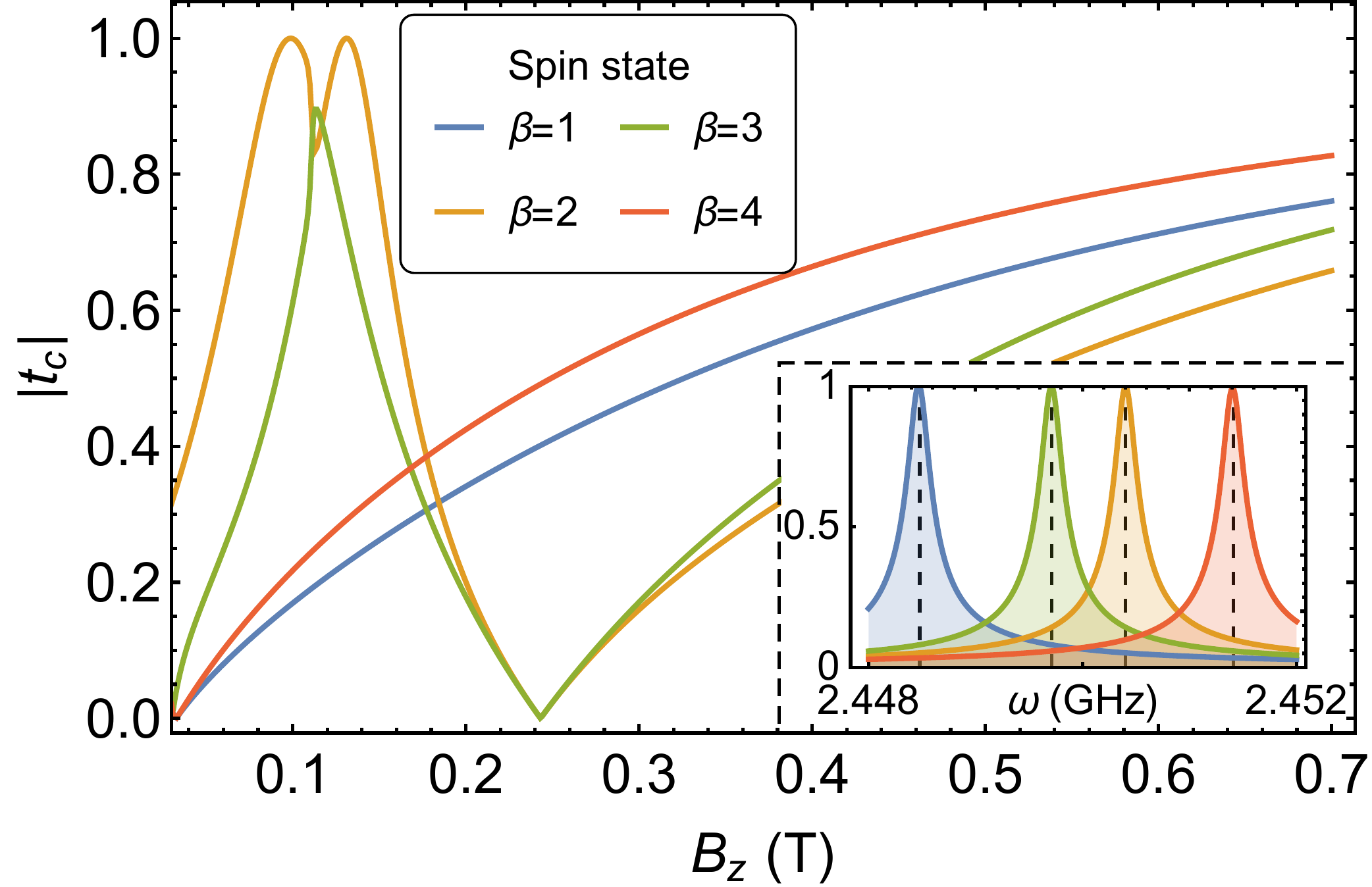}
    \caption{Absolute value of the transmission for $\omega=\Omega$ as a function of the longitudinal field. We have considered a cavity with dissipative rates $\gamma_i=10^{-4}$, which are of the order of $T_2$ for the molecule. The inset shows the broadening of the transmission peaks as a function of $\omega$ for $B_z=0.02$T.}
    \label{fig:Fig6}
\end{figure}

It is also important to discuss the role of the cavity $Q$-factor in the readout. So far, we have assumed that spins decohere much faster than cavity photons, as it is usually the case. This produces sharp peaks in the transmission at specific frequencies, which is one of the reasons why phase measurements are more robust to imperfections during the readout. This picture can change if the decoherence rate of the cavity, $\gamma$, is close to that of the spins, $\eta$. In this case, the transmission peaks associated with different spin states broaden and partially overlap (see Fig.~\ref{fig:Fig6}, inset). At any frequency close to one of these resonances, this gives rise to a continuous change in the transmission and produces the interesting behaviour shown in Fig.~\ref{fig:Fig6}, as a function of magnetic field. In this case, one can see that measuring the transmission at a single frequency $\omega=\Omega$ and for a single magnetic field value gives a different, an nonzero outcome for each spin state. Eventually, this could allow performing a single shot readout of the spin state provided that the transmission differences can be experimentally resolved. 
Obviously, a less coherent cavity also restricts the effective lifetime of the qudit state to which it is coupled. Yet, this coherence loss could be compensated by the reduction in the number of measurements required.

As we discussed in the case of GdW$_{30}$, all this applies to the case in which the spin line widths are dominated by the homogeneous broadening. Synthesizing single crystals hosting molecular dimers, such as CeEr, diluted in a diamagnetic host is still very challenging, 
whereas the use of frozen solutions leads to very broad resonances on account of the random distribution of anisotropy axes. This underlines the importance of using realistic calculations to guide the choice, and the design, of suitable candidates. A promising one is discussed in the next section.
\subsection{$^{173}\text{Yb-trensal}$: an electronuclear spin qudit}
Experimentally, it has been shown that a single spin-photon interaction, of the order of kHz, can be achieved in constrained resonators~\cite{Gimeno2020}, and that there are already molecular spins showing ${\rm T}_{2}$ values in the vicinity of $1$ms~\cite{Zadrozny2015a}.
However, as readout will likely be first performed in spin ensembles, the limiting factor will be the 
inhomogeneous broadening rather than $T_2$, which will be difficult to reduce in the previous cases of $\text{GdW}_{30}$ and $\text{[CeEr]}$. For this reason we now consider the case of $^{173}\text{Yb-trensal}$, which combines a nuclear spin qudit ($I=5/2$) coupled to an effective $S=1/2$ electronic spin doublet.

In contrast with the previous examples, the photon coupling to the nuclear spin is much weaker than that to the electron spin, due to the small ratio between the nuclear and the electronic magneton, $\mu_N/\mu_B \ll1$.
However, the presence of a strong hyperfine interaction, characteristic of lanthanide ions, mediates the indirect coupling between the nuclear spin and the photons, and therefore, the manipulation of the all the spin levels defining the qubit. 
This means that in principle, it should be possible to perform the 
dispersive readout of all electronuclear spin states, including states differing only by its nuclear spin 
projection $m_{I}$. We now show that this is possible in practice.
In addition, this molecular system has been considered as a good candidate for the implementation of quantum error correction codes~\cite{Hussain2018,Chiesa2020}, in which the nuclear spin helps to define the computational basis of the logical qubit and the electronic spin is exploited to detect errors. 
With this idea in mind, we now discuss how to readout the logical qubit states, and demonstrate that due to the QND nature of the measurement, we could also detect errors to neighboring energy levels at any arbitrary time of the error correction protocol.

Let us first introduce the Hamiltonian for the $^{173}\text{Yb-trensal}$ molecule, which is given by:
\begin{align}
    \mathcal{H}_S=&\mu_{B}\vec{B}\cdot\hat{g}\cdot\vec{S}+\mu_{N}g_{I}\vec{B}\cdot\vec{I}+pI_{z}^{2}\\\nonumber
    &+A_{\parallel}S_{z}I_{z}+A_{\perp}\left(S_{x}I_{x}+S_{y}I_{y}\right),
\end{align}
where the first line contains the coupling of the electronic and nuclear spin to the external magnetic field, $\mu_N$ is the nuclear magneton, $g_I$ is the nuclear g-factor, $p$ describes the nuclear quadrupolar interaction and the second line contains the anisotropic hyperfine interaction.
From the experimental fitting of the parameters one finds $\hat{g}=(g_\perp,g_\perp,g_\parallel)$, $g_{\perp}=2.935$, $g_{\parallel}=4.225$, $g_I=-0.02592$, $A_\parallel=-0.897$GHz, $A_\perp=-0.615$GHz and $p=-0.066$GHz.
\begin{figure}
    \includegraphics[width=1.0\columnwidth]{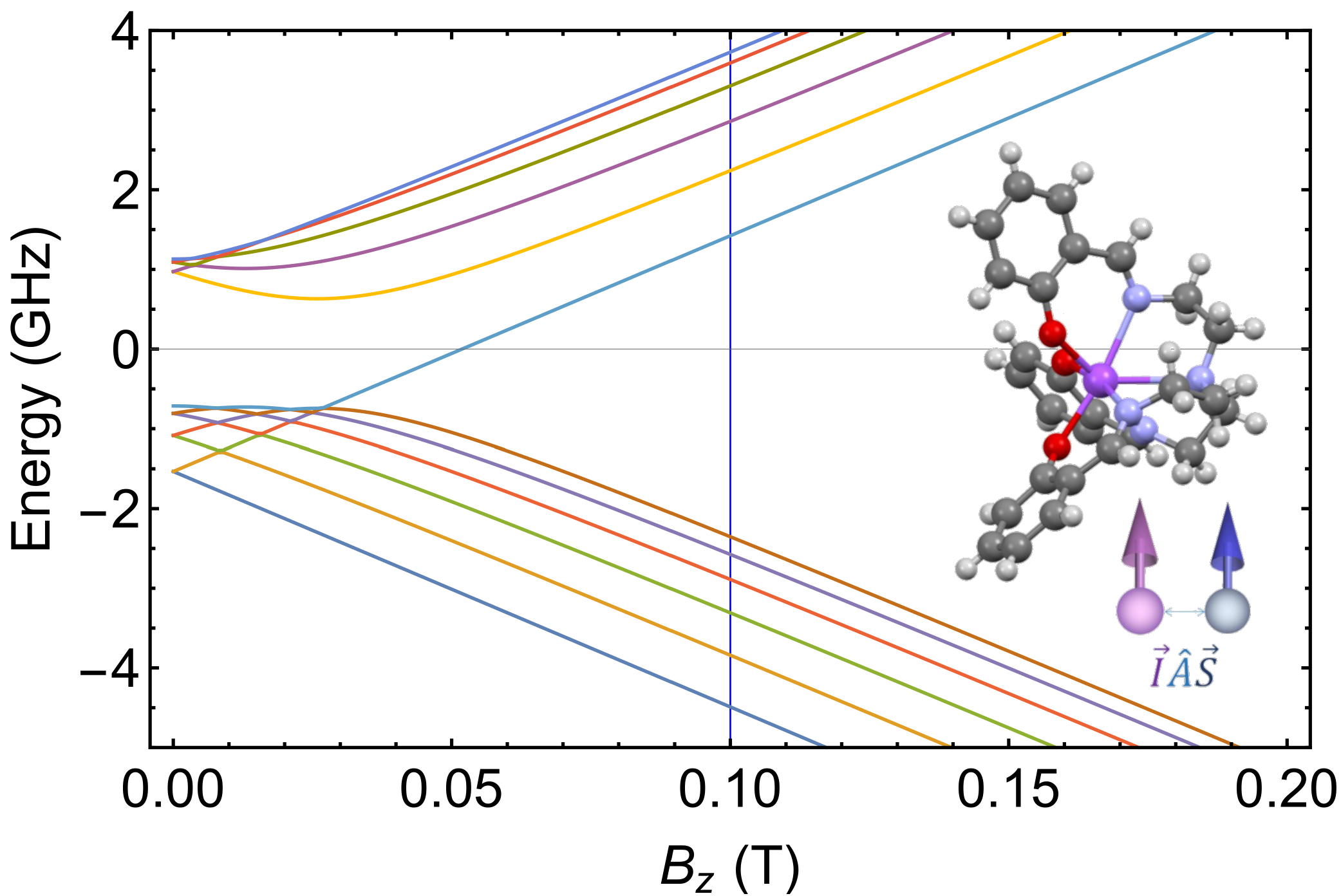}
    \caption{Spectrum of the $^{173}\text{Yb-trensal}$ molecule as a function of $B_z$. The vertical (blue) line indicates the field chosen to perform the readout.}
    \label{fig:Spectrum-Trensal}
\end{figure}
In Fig.~\ref{fig:Spectrum-Trensal} we plot the spectrum of the $^{173}\text{Yb-trensal}$ molecule, as a function of the longitudinal field. It shows how the magnetic field controls the hybridization of the electronuclear states.

Regarding the spin-photon coupling, the $\Lambda_{\vec{\alpha}}$ tensor now takes a different form than in the previous cases, due to the presence of the nuclear spin. 
Concretely, the interaction Hamiltonian via the Zeeman term is given by:
\begin{equation}
    \mathcal{H}_I = \left(a^{\dagger}+a\right)\left(\vec{\lambda}_{S}\cdot\hat{g}_{S}\cdot\vec{S}+\vec{\lambda}_{I}\cdot\hat{g}_{I}\cdot\vec{I}\right),\ \label{eq:TrensalV}
\end{equation}
Although we will consider all terms in Eq.~\eqref{eq:TrensalV}, it is also a good approximation to neglect the contribution proportional to the nuclear magneton $\vec{\lambda}_I$.

For isotopically pure crystals of $^{173}\text{Yb-trensal}$ with a $1\%$ concentration, it is possible to estimate, from experimental measurements~\cite{rollano2022}, a collective spin-photon coupling of the order of $\lambda_{S}^{x}=20$MHz, which is enough to achieve coherent coupling. Simultaneously, the inhomogeneous broadening is of the order of $\eta=12$MHz for this concentration.

\begin{figure}
    \centering
    \includegraphics[width=1.0\columnwidth]{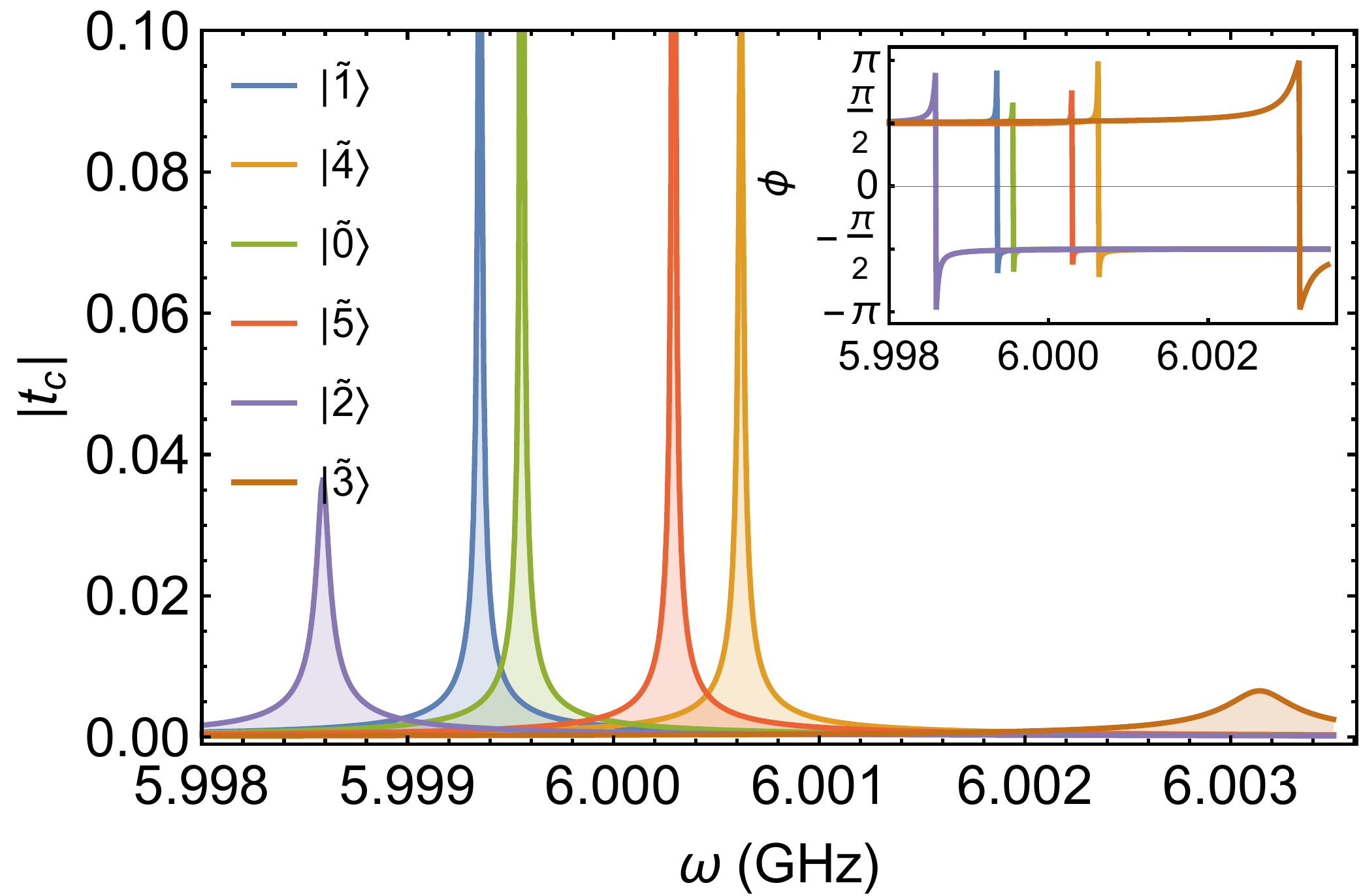}
    \caption{Resonator transmission for the ground state multiplet at $B_z=0.1$T and $\Omega=6$GHz. The states $|\tilde{1}\rangle$ and $|\tilde{4}\rangle$ encode the logical qubits with quantum error correction, but the other states can also be detected. The inset shows the phase measurement, which is more robust for the case of damped transmission peaks, as for state $|\tilde{3}\rangle$. Parameters: $\gamma=10^{-3}$MHz, $\eta=12$MHz and collective spin-photon coupling $\lambda_{S}^{x}=20$MHz.}
    \label{fig:Trensal-transmission}
\end{figure}

The use of $^{173}\text{Yb-trensal}$ as a qubit with embedded error correction exploits the six levels multiplet associated with the electronic spin projection $m_{S}=-1/2$, which allows to implement a minimal code for protection against single amplitude or phase errors~\cite{Pirandola2008}. For this, one defines the logical qubit in terms of the states $|\tilde{1}\rangle=|-\frac{1}{2},{-\frac{3}{2}}\rangle$ and $|\tilde{4}\rangle=|-\frac{1}{2},{\frac{3}{2}}\rangle$, while all the other states in the multiplet are auxiliary to implement the code.
Therefore, we are primarily interested in reading out states $|\tilde{1}\rangle$ and $|\tilde{4}\rangle$, to know the state of the logical qubit.
Nevertheless, our readout protocol should also allow us to detect the other states in the multiplet, in case that we want to study the inner workings of the error correction code or the influence of more complex errors, produced e.g., by correlated errors with other spins~\cite{Prokof_ev_2000,GomezLeon-2019}.
We show in Fig.~\ref{fig:Trensal-transmission} the transmission calculated for the different qudit states 
and the experimental parameters.
Importantly, the logical qubit peaks are well separated, indicating that the qubit can be measured using the dispersive readout. This also implies that, although the level broadening is of the order of $12$MHz, a high resonator $Q$-factor is more important. Actually, the dominant effect of the spectral broadening is to reduce the height of the peaks, which can be mitigated by performing phase measurements. Furthermore, the other states in the multiplet can also be measured, allowing to detect and correct errors.
\section{Conclusions}
We have presented a general framework to study the transmission of a superconducting cavity coupled to magnetic molecules of arbitrary spin in the dispersive regime. Our framework incorporates their multi-level structure and the influence of the magnetic anisotropy, which is ubiquitous in these molecules. It is valid in both, the strong and the weak coupling regimes of spin-photon interaction. Besides, it can be adapted to more complex situations involving several coupled spins within each molecule. We have shown that the transmission depends on the spin state in a way that allows a uni-vocal determination of this state. Our results provide a generalisation of the well-known dispersive readout of qubits, which is broadly used in different quantum computing schemes. Therefore, molecular spin qudits can not only be coherently controlled to implement different quantum gates or protocols, but the results can also be readout with technologically available methods.

The multilevel structure of the qudit introduces however some changes to the readout protocol: $d-1$ measurement pulses with judiciously-chosen frequencies, instead of just one as in the case of qubits, are required to determine the spin state. Besides this method, we have shown that a single shot readout protocol is also feasible, provided that the bare resonator line-width be comparable to the decoherence rate of the spin qudit.

In addition, there are qualitatively additional features. In general, the magnetic anisotropy, which leads to the nonlinear
arrangement of the spin qudit levels, affects the non-demolition nature of the dispersive readout. This property can, however, be restored by an appropriate control of the magnetic fields coupled to the molecule. We have also developed an
effective Hamiltonian to describe the coupled
qudit-cavity system that generalises the well-known effective spin-photon interaction  Hamiltonian for $S=1/2$ to arbitrary spin with non-linear magnetic anisotropy terms. This formalism can also be helpful to study the state dynamics, for the implementation of quantum gates and protocols~\cite{GomezLeon2022}. It also provides a theoretical framework to analyze other applications of the cavity-spin coupling in the dispersive regime, e.g. to perform magnetic spectroscopy measurements~\cite{Bonizzoni2021}.

As an application and illustration of our method, we have considered three relevant examples of molecular spin qudits currently
under study. The first is a Gd-based single ion magnet with $S=7/2$ which encodes three qubits and that was used to implement a
Toffoli gate. The second is a molecular dimer {[}CeEr{]} in which it is possible to define a fully addressable two-qubit system
and where a CNOT gate has been implemented. The third corresponds to the molecule of $^{173}\text{Yb-trensal}$ with $d=12$ electronuclear spin states, where error correction can be implemented. 
We have shown that the cavity transmission can be used to readout the spin qudit states, although the inhomogenous broadening might limit the application of this method in the first two systems. For Yb-trensal, we have shown that coupling of cavity photons to electronic spin transitions, in the range of a few GHz, 
allows reading out the nuclear spin states. This effect results from the strong hyperfine interaction that characterizes lanthanide spin qubits, and might considerably simplify the implementation of proof-of-concept experiments in this and related systems. The theory will likely provide a useful guide for the design of suitable molecular spin qudits, which should combine a proper level anharmonicity, a not too large distribution of resonant frequencies to make frequency shifts associated with all states observable, and good material properties (e.g. a sufficiently low inhomogeneous broadening).

\begin{acknowledgments}
We acknowledge the funding from the Spanish MICIN grants 
PGC2018-094792-B-I00, RTI2018-096075-B-C21 and PCI2018-093116 (MCIU/AEI/FEDER, UE), the European Union's
Horizon 2020 research and innovation programme (QUANTERA project SUMO, FET-OPEN grant 862893 FATMOLS), the Gobierno de
Arag\'{o}n grant E09-17R-Q-MAD and the CSIC Quantum Technology Platform PT-001.
\end{acknowledgments}

\bibliography{main.bib}

\begin{widetext}

\appendix

\section{Cavity transmission for an ensemble of magnetic molecules}
\label{app:ensemble}
Here we derive the cavity transmission for the case of an ensemble of molecules, using the equation of motion technique. We start from the full Hamiltonian:
\begin{equation}
    \mathcal{H}= \mathcal{H}_B + \mathcal{H}_c + \mathcal{H}_S + \mathcal{H}_I
\end{equation}
where $\mathcal{H}_S=\sum_{i=1}^N \sum_{\alpha=1}^{2S+1} E_{\alpha} X_i^{\alpha,\alpha}$, $\mathcal{H}_I=(a^\dagger+a)\sum_{i=1}^N\sum_{\vec{\alpha=1}}^{2S+1} \Lambda_{i}^{\vec\alpha}X_i^{\vec{\alpha}}$, $\mathcal{H}_c=\Omega a^\dagger a$ and $\mathcal{H}_B$ is the Hamiltonian for the modes in the transmission line, which can be written as:
\begin{equation}
   \mathcal{H}_B = \sum_{l}\int_{-\infty}^{\infty}\omega b_{\omega,l}^{\dagger}b_{\omega,l}d\omega+i\sum_{l}\int_{-\infty}^{\infty}d\omega\left[\kappa_{l}\left(\omega\right)b_{\omega,l}^{\dagger}a-\kappa_{l}^{\ast}\left(\omega\right)a^{\dagger}b_{\omega,l}\right]
\end{equation}
being $b_{\omega,l}$ the photon operator at energy $\omega$ and port $l$, and $\kappa_l(\omega)=\sqrt{\frac{\gamma_l}{2\pi}}$ under the First Markov approximation. To integrate-out the transmission line modes, we calculate of the Heisenberg equation of motion for the transmission line:
\begin{equation}
    \partial_{t}b_{\omega,l}\left(t\right)=-i\omega b_{\omega,l}\left(t\right)+\sqrt{\frac{\gamma_{l}}{2\pi}}a\left(t\right)
\end{equation}
which can be directly integrated to give:
\begin{equation}
    b_{\omega,l}\left(t\right)=b_{\omega,l}\left(t_{0}\right)e^{-i\omega\left(t-t_{0}\right)}+\sqrt{\frac{\gamma_{l}}{2\pi}}\int_{t_{0}}^{t}e^{-i\omega\left(t-\tau\right)}a\left(\tau\right)d\tau
\end{equation}
and leads to the standard Langevin equation for the cavity modes (we define the total dissipation rate $\gamma=\sum_{l}\gamma_{l}$):
\begin{equation}
    \partial_{t}a\left(t\right)=-i\left(\Omega-i\frac{\gamma}{2}\right)a\left(t\right)-i\sum_{i=1}^{N}\sum_{\vec{\alpha}=1}^{2S+1}\Lambda_{i}^{\vec{\alpha}}X_{i}^{\vec{\alpha}}\left(t\right)-\sum_{l}\sqrt{\gamma_{l}}b_{\text{in},l}\left(t\right)
\end{equation}
with the standard definition for the input modes:
\begin{equation}
    b_{\text{in},l}\left(t\right)=\frac{1}{\sqrt{2\pi}}\int_{-\infty}^{\infty}d\omega e^{-i\omega\left(t-t_{0}\right)}b_{\omega,l}\left(t_{0}\right)
\end{equation}
To close the system of equations we calculate the Heisenberg equation of motion for an arbitrary Hubbard operator:
\begin{equation}
    \partial_{t}X_{i}^{\vec{\alpha}}(t)=iE_{\vec{\alpha}}X_{i}^{\vec{\alpha}}(t)+i\left(a(t)+a^{\dagger}(t)\right)\sum_{\mu}\left(\Lambda_{i}^{\mu,\alpha_{1}}X_{i}^{\mu,\alpha_{2}}(t)-\Lambda_{i}^{\alpha_{2},\mu}X_{i}^{\alpha_{1},\mu}(t)\right)
\end{equation}
which requires a truncation scheme to close the system of equations (because it couples to additional many-body operators such as $aX^{\alpha_1,\mu}$). 
In addition, notice that we have assumed that the energies of the spins are all equal, because their chemical synthesis produces identical molecules and the external magnetic field is homogeneous.
However, this could not be the case due to local substrate effects or if the system under consideration is not made of identical qudits.
In that case the calculation just requires to also define a continuous spectral density to solve the coupled equations of motion.

As experiments typically display weak coupling between individual spins and the cavity, one can write the following decomposition, to separate the photonic and the spin part of the many-body operators:
\begin{eqnarray}
    aX^{\mu,\nu} &=& a\langle X^{\mu,\nu} \rangle+\langle a \rangle X^{\mu,\nu}+\delta a \delta X^{\mu,\nu}\nonumber\\
    &\simeq& a\langle X^{\mu,\nu} \rangle \delta_{\mu,\nu}
\end{eqnarray}
This approximation is only valid for weak interactions, but its valid to describe the weak and strong coupling regimes~\cite{Forn-Diaz2019}. The reason is that, in this limit, fluctuations $\delta a \delta X^{\mu,\nu}$ are very small (this can be argued in terms of a mean field description), and that off-diagonal averages are also small compared with diagonal ones (notice that the Hamiltonian is completely diagonal in absence of coupling). This allows us to neglect the averages $\langle a \rangle$, $\langle a^\dagger \rangle$ and $\langle X^{\mu,\nu} \rangle$ for $\mu\neq\nu$. In addition, we will neglect the term proportional to $a^\dagger \langle X^{\mu,\nu} \rangle$, because its correction is of higher order. This is not strictly necessary to close the set of equations, but allows to obtain more compact expressions. 
In summary, after a Fourier transform, the solution for the equation of motion of an arbitrary Hubbard operator takes the following form:
\begin{equation}
    X_{i}^{\vec{\alpha}}(\omega) \simeq-a\frac{\Lambda_{i}^{\alpha_{2},\alpha_{1}}}{\omega+E_{\vec{\alpha}}}\left(\langle X_{i}^{\alpha_{2},\alpha_{2}}\rangle-\langle X_{i}^{\alpha_{1},\alpha_{1}}\rangle\right)
\end{equation}
Finally, as a last assumption to obtain a compact expression for the transmission, we consider the system in a diagonal density matrix in the basis of Hubbard operators $\rho=\sum_{\alpha=1}^{2S+1} p_\alpha X^{\alpha,\alpha}$, for all the different molecules in the ensemble. This is adequate due to the state preparation carried out in the experimental setups. All these assumptions result in the final expression for the cavity mode:
\begin{equation}
    a\left(\omega\right)=\frac{i\sum_{l}\sqrt{\gamma_{l}}b_{\text{in},l}(\omega)}{\Omega-\omega-i\frac{\gamma}{2}+\sum_{i=1}^{N}\sum_{\vec{\alpha}=1}^{2S+1}\frac{p_{\vec{\alpha}}\left|\Lambda_{i}^{\vec{\alpha}}\right|^{2}}{\omega+E_{\vec{\alpha}}+i\eta}}
\end{equation}
where $|\Lambda_{i}^{\vec{\alpha}}|^{2}=\Lambda_{i}^{\alpha_1, \alpha_2}\Lambda_{i}^{\alpha_2, \alpha_1}$, $E_{\vec{\alpha}}=E_{\alpha_1}-E_{\alpha_2}$, $p_{\vec{\alpha}}=p_{\alpha_1} - p_{\alpha_2} $ and $\eta$ is the phenomenological spectral broadening of the spin energy levels. To obtain the cavity transmission one just needs to make use of the standard input-output relation in $t_c=\langle b_{\text{out},2}\rangle / \langle b_{\text{in},1} \rangle$. 
This result demonstrates that the frequency shift produced by the interaction between the cavity mode and the ensemble of molecules is enhanced a factor $N$ with respect to the case of a single molecule.
Moreover, although the derivation requires to make an assumptions about the coupling strength between a single spin and the cavity photons in the dispersive regime (i.e., for an off-resonant condition between spin transitions and the cavity), it does not require a condition with respect to $\gamma$ or $\eta$, which allows to explore both, the weak and the strong coupling regimes. 
In addition, if one is interested in going beyond the strong coupling regime, the formula remains valid by just considering a mean field basis~\cite{Perez-Gonzalez2021}.

To explicitly see the enhancement in the cavity frequency shit, assume that all the molecules are prepared in state $\beta$ (i.e., $p_{\alpha}=1$ for $\alpha=\beta$, otherwise $p_{\alpha}=0$). Then, the sum in the denominator of the transmission contains only two terms:
\begin{equation}
        t_c\left(\omega\right)=\frac{i\sqrt{\gamma_{1}\gamma_{2}}}{\Omega-\omega-i\frac{\gamma}{2}+\sum_{i=1}^{N}\sum_{\alpha=1}^{2S+1}\left(\frac{\Lambda_{i}^{\beta,\alpha}\Lambda_{i}^{\alpha,\beta}}{\omega+E_{\beta}-E_{\alpha}+i\eta}-\frac{\Lambda_{i}^{\alpha,\beta}\Lambda_{i}^{\beta,\alpha}}{\omega+E_{\alpha}-E_{\beta}+i\eta}\right)}
\end{equation}
Re-organizing the denominator we can write:\begin{equation}
        t_c\left(\omega\right)=\frac{i\sqrt{\gamma_{1}\gamma_{2}}}{\Omega-\omega-i\frac{\gamma}{2}+\sum_{i=1}^{N}\sum_{\alpha=1}^{2S+1}\frac{2\left|\Lambda_{i}^{\alpha,\beta}\right|^{2}\left(E_{\alpha}-E_{\beta}\right)}{\left(\omega+i\eta\right)^{2}-\left(E_{\alpha}-E_{\beta}\right)^{2}}}
\end{equation}
where it is clear that the shift in the frequency photon, measured at $\omega$ is:
\begin{equation}
    \delta\tilde{\Omega}_{\beta}\left(\omega\right)=\sum_{i=1}^{N}\sum_{\alpha=1}^{2S+1}\frac{2\left|\Lambda_{i}^{\alpha,\beta}\right|^{2}E_{\alpha,\beta}}{\left(\omega+i\eta\right)^{2}-E_{\alpha,\beta}^{2}}
\end{equation}
\section{Derivation of the effective Hamiltonian and check for the qubit case}
\label{app:2ls}

The calculation of the S-W transformation requires to first find the ansatz for the transformation $\mathcal{S}$. For this, it is common to find is operator form from the commutator $[\mathcal{H}_0,\mathcal{H}_I]$. It results in the following expression for the transformation:
\begin{equation}
\mathcal{S}=\sum_{\vec{\beta}=1}^{2S+1}\left(\Gamma_{+}^{\vec{\beta}}a^{\dagger}+\Gamma_{-}^{\vec{\beta}}a\right)X^{\vec{\beta}}, 
\end{equation}
with $\Gamma_{\pm}^{\vec{\beta}}=\frac{\Lambda_{\vec{\beta}}}{E_{\vec{\beta}}\pm\Omega}$.
The calculation of the effective Hamiltonian to second order in $\Lambda_{\vec{\alpha}}$: $\tilde{\mathcal{H}}\simeq \mathcal{H}_0+\frac{1}{2}\left[ \mathcal{S},\mathcal{H}_I \right]$, is obtained from the calculation of the commutator $\left[ \mathcal{S},\mathcal{H}_I \right]$, which yields:
\begin{eqnarray}
\left[S,\mathcal{H}_{I}\right] & = & 2a^{\dagger}a\sum_{\beta_{i},\alpha=1}^{2S+1}\Lambda_{\beta_{1},\alpha}\Lambda_{\alpha,\beta_{2}}\left(\frac{E_{\beta_{1},\alpha}}{E_{\beta_{1},\alpha}^{2}-\Omega^{2}}-\frac{E_{\alpha,\beta_{2}}}{E_{\alpha,\beta_{2}}^{2}-\Omega^{2}}\right)X^{\vec{\beta}}\nonumber\\
 &  & +\sum_{\beta_{i},\alpha=1}^{2S+1}\Lambda_{\beta_{1},\alpha}\Lambda_{\alpha,\beta_{2}}\left(\frac{1}{E_{\beta_{1},\alpha}-\Omega}-\frac{1}{E_{\alpha,\beta_{2}}+\Omega}\right)X^{\vec{\beta}}\nonumber\\
 &  & +\left(a^{\dagger}\right)^{2}\sum_{\beta_{i},\alpha=1}^{2S+1}\Lambda_{\beta_{1},\alpha}\Lambda_{\alpha,\beta_{2}}\left(\frac{1}{E_{\beta_{1},\alpha}+\Omega}-\frac{1}{E_{\alpha,\beta_{2}}+\Omega}\right)X^{\vec{\beta}}\nonumber\\
 &  & +a^{2}\sum_{\beta_{i},\alpha=1}^{2S+1}\Lambda_{\beta_{1},\alpha}\Lambda_{\alpha,\beta_{2}}\left(\frac{1}{E_{\beta_{1},\alpha}-\Omega}-\frac{1}{E_{\alpha,\beta_{2}}-\Omega}\right)X^{\vec{\beta}}
\end{eqnarray}
The final form of the effective Hamiltonian is obtained by adding the unperturbed terms, resulting in the following expression:
\begin{eqnarray}
\tilde{\mathcal{H}} & \simeq & \sum_{\alpha}E_{\alpha}X^{\alpha,\alpha}+\frac{1}{2}\sum_{\beta_{i},\alpha=1}^{2S+1}\Lambda_{\beta_{1},\alpha}\Lambda_{\alpha,\beta_{2}}\left(\frac{1}{E_{\beta_{1},\alpha}-\Omega}+\frac{1}{E_{\beta_{2},\alpha}-\Omega}\right)X^{\vec{\beta}}\nonumber \\
 &  & +\Omega a^{\dagger}a+a^{\dagger}a\sum_{\beta_{i},\alpha=1}^{2S+1}\Lambda_{\beta_{1},\alpha}\Lambda_{\alpha,\beta_{2}}\left(\frac{E_{\beta_{1},\alpha}}{E_{\beta_{1},\alpha}^{2}-\Omega^{2}}+\frac{E_{\beta_{2},\alpha}}{E_{\beta_{2},\alpha}^{2}-\Omega^{2}}\right)X^{\vec{\beta}}\nonumber \\
 &  & +\frac{1}{2}\left(a^{\dagger}\right)^{2}\sum_{\beta_{i},\alpha=1}^{2S+1}\Lambda_{\beta_{1},\alpha}\Lambda_{\alpha,\beta_{2}}\left(\frac{1}{E_{\beta_{1},\alpha}+\Omega}+\frac{1}{E_{\beta_{2},\alpha}-\Omega}\right)X^{\vec{\beta}}\nonumber \\
 &  & +\frac{1}{2}a^{2}\sum_{\beta_{i},\alpha=1}^{2S+1}\Lambda_{\beta_{1},\alpha}\Lambda_{\alpha,\beta_{2}}\left(\frac{1}{E_{\beta_{1},\alpha}-\Omega}+\frac{1}{E_{\beta_{2},\alpha}+\Omega}\right)X^{\vec{\beta}}\label{eq:Full-effective-H}
\end{eqnarray}
This Hamiltonian is valid up to second order in $\Lambda_{\vec{\alpha}}$ and has a large number of contributions, including second order photon processes. Nevertheless, a simpler form can be obtained if we consider that diagonal averages $\langle X^{\alpha,\alpha}\rangle$ and $\langle a^{\dagger}a\rangle$ dominate over off-diagonal ones. This is a good approximation if the artificial molecule is weakly perturbed by the interaction with the cavity (however, if one is interested in the time-evolution of the system, it might be important to include the off-diagonal terms to capture the long-time behavior). Therefore, we can write the final form of the Hamiltonian used in the main text as:
\begin{eqnarray}
\tilde{\mathcal{H}} & \simeq & \sum_{\alpha}E_{\alpha}X^{\alpha,\alpha}+\sum_{\alpha,\beta=1}^{2S+1}\frac{\left|\Lambda_{\alpha,\beta}\right|^{2}}{E_{\alpha,\beta}-\Omega}X^{\alpha,\alpha}\nonumber \\
 &  & +a^{\dagger}a\left(\Omega+2\sum_{\alpha,\beta=1}^{2S+1}\frac{E_{\alpha,\beta}\left|\Lambda_{\alpha,\beta}\right|^{2}}{E_{\alpha,\beta}^{2}-\Omega^{2}}X^{\alpha,\alpha}\right)
\end{eqnarray}
where we have defined $\left|\Lambda_{\alpha,\beta}\right|^{2}=\Lambda_{\alpha,\beta}\Lambda_{\beta,\alpha}$.

As a check, we can reproduce the case of several qubits interacting with a single photonic mode in a cavity:
\begin{equation}
    H = \sum_{i}\frac{\Delta_{i}}{2}\sigma_{i}^{z}+\Omega a^{\dagger}a+\left(a^{\dagger}+a\right)\sum_{i}g_{i}\sigma_{i}^{x}
    = \sum_{i}\frac{\Delta_{i}}{2}\left(X_{i}^{+,+}-X_{i}^{-,-}\right)+\Omega a^{\dagger}a+\left(a^{\dagger}+a\right)\sum_{i}g_{i}\left(X_{i}^{+,-}+X_{i}^{-,+}\right)
\end{equation}
Inserting these couplings in the expression for the effective Hamiltonian we find that the S-W transformation is given by:
\begin{equation}
    \mathcal{S}=\sum_{j}g_{j}\left(\frac{a^{\dagger}X_{j}^{+,-}-aX_{j}^{-,+}}{\Delta_{j}+\Omega}+\frac{aX_{j}^{+,-}-a^{\dagger}X_{j}^{-,+}}{\Delta_{j}-\Omega}\right)
\end{equation}
and the effective Hamiltonian by:
\begin{eqnarray}
\tilde{\mathcal{H}}&=&\sum_{i}\frac{\Delta_{i}}{2}\left(X_{i}^{+,+}-X_{i}^{-,-}\right)+\sum_{i}g_{i}^{2}\left(\frac{X_{i}^{+,+}}{\Delta_{i}-\Omega}-\frac{X_{i}^{-,-}}{\Delta_{i}+\Omega}\right)\nonumber\\
&&+a^{\dagger}a\left[\Omega+\sum_{i}\frac{2\Delta_{i}g_{i}^{2}}{\Delta_{i}^{2}-\Omega^{2}}\left(X_{i}^{+,+}-X_{i}^{-,-}\right)\right]\nonumber\\
&&+\left(a^{\dagger}a^{\dagger}+aa\right)\sum_{i}\frac{\Delta_{i}g_{i}^{2}}{\Delta_{i}^{2}-\Omega^{2}}\left(X_{i}^{+,+}-X_{i}^{-,-}\right)\nonumber\\
&&+\sum_{i,j\neq i}\frac{\Omega g_{i}g_{j}}{\Delta_{j}^{2}-\Omega^{2}}\left(X_{i}^{+,-}+X_{i}^{-,+}\right)\left(X_{j}^{+,-}+X_{j}^{-,+}\right)
\end{eqnarray}
Importantly, notice how neglecting off-diagonal contributions, we eliminate the quadratic photon terms $a^2$ and $(a^{\dagger})^2$, but also the effective qubit-qubit interaction commonly used for quantum gates engineering. This is unimportant for the current case of quantum spectroscopy, where we are interested in the readout of the qubits state. The main reason is that readout is a fast process and the qubits do not have enough time to entangle via the effective interaction, which takes a time of the order $\tau \sim g_i^{-2}$. However, if one is interested in effective interactions or these time-scales, it will be important to keep off-diagonal terms as well~\cite{GomezLeon-2019,GomezLeon2022}.

\section{Non-demolition measurement}
\label{app:nondem}
The calculation of the commutator between the unperturbed spin Hamiltonian
and the photon frequency shift term results in:
\begin{equation}
\left[\mathcal{H}_{S},\tilde{\mathcal{V}}\right]=\sum_{\beta_{i}=1}^{2S+1}E_{\vec{\beta}}\Phi_{\vec{\beta}}X^{\vec{\beta}}
\end{equation}
where
\begin{equation}
\tilde{\mathcal{V}}=\sum_{\beta_{i},\alpha=1}^{2S+1}\Lambda_{\beta_{1},\alpha}\Lambda_{\alpha,\beta_{2}}\left(\frac{E_{\beta_{1},\alpha}}{E_{\beta_{1},\alpha}^{2}-\Omega^{2}}+\frac{E_{\beta_{2},\alpha}}{E_{\beta_{2},\alpha}^{2}-\Omega^{2}}\right)X^{\vec{\beta}}
\end{equation}
is the effective spin-photon interaction in Eq.~\eqref{eq:Full-effective-H}
and
\begin{equation}
\Phi_{\vec{\beta}}=\sum_{\alpha=1}^{2S+1}\Lambda_{\beta_{1},\alpha}\Lambda_{\alpha,\beta_{2}}\left(\frac{E_{\beta_{1},\alpha}}{E_{\beta_{1},\alpha}^{2}-\Omega^{2}}+\frac{E_{\beta_{2},\alpha}}{E_{\beta_{2},\alpha}^{2}-\Omega^{2}}\right).
\end{equation}
This expression is valid up to second order in $\Lambda_{\vec{\beta}}$
and indicates that in general the readout will not be a non-demolition
measurement. To analyze in detail this result we can consider the
qubit case:
\begin{equation}
\sum_{\beta_{i}=1}^{2S+1}E_{\vec{\beta}}\Phi_{\vec{\beta}}X^{\vec{\beta}}=E_{+,-}^{2}\frac{\Lambda_{-,-}-\Lambda_{+,+}}{E_{+,-}^{2}-\Omega^{2}}\left(\Lambda_{+,-}X^{+,-}-\Lambda_{-,+}X^{-,+}\right)
\end{equation}
which indicates that only if the spin-photon interaction is purely
transversal (i.e., if $\Lambda_{\alpha,\alpha}=0$), the readout is
a non-demolition measurement. Crucially, these extra terms can be compensated by just rotating the unperturbed part of the Hamiltonian in such a way that the interaction term becomes purely transverse to the qubit quantization axis.

\section{Molecule with spin $S=1$}

We consider a $S=1$ molecule with uniaxial anisotropy and Hamiltonian:
\begin{equation}
\mathcal{H}=D\left(S^{z}\right)^{2}+\xi_{z}S^{z}+\Omega a^{\dagger}a+\lambda_x g_{x}\left(a^{\dagger}+a\right)S^{x}
\end{equation}
From the Hamiltonian, we can obtain the relevant unperturbed energies $E_\pm=D\pm \xi_z$ and $E_0 =0$, and the elements of the interaction tensor:
\begin{equation}
\Lambda_{\vec{\alpha}}=\frac{\lambda_{x}g_{x}}{2}\sqrt{2-\alpha_{1}\alpha_{2}}\left(\delta_{\alpha_{1},\alpha_{2}+1}+\delta_{\alpha_{2},\alpha_{1}+1}\right)
\end{equation}
Then, inserting this values in Eq.~\eqref{eq:Effective-H}, we can calculate the full effective Hamiltonian:
\begin{eqnarray}
\tilde{\mathcal{H}}&\simeq&\Omega a^{\dagger}a-4\left(\frac{\lambda_{x}g_{x}}{2}\right)^{2}\left[\frac{D+\Omega}{\left(D+\Omega\right)^{2}-\xi_{z}^{2}}+a^{\dagger}a\left(\frac{D-\xi_{z}}{\left(D-\xi_{z}\right)^{2}-\Omega^{2}}+\frac{D+\xi_{z}}{\left(D+\xi_{z}\right)^{2}-\Omega^{2}}\right)\right]X^{0,0}\nonumber\\
&&+\left[D+\xi_{z}+\left(\frac{\lambda_{x}g_{x}}{2}\right)^{2}\left(\frac{2}{D+\xi_{z}-\Omega}+a^{\dagger}a\frac{4\left(D+\xi_{z}\right)}{\left(D+\xi_{z}\right)^{2}-\Omega^{2}}\right)\right]X^{+,+}\nonumber\\
&&+\left[D-\xi_{z}+\left(\frac{\lambda_{x}g_{x}}{2}\right)^{2}\left(\frac{2}{D-\xi_{z}-\Omega}+a^{\dagger}a\frac{4\left(D-\xi_{z}\right)}{\left(D-\xi_{z}\right)^{2}-\Omega^{2}}\right)\right]X^{-,-}\nonumber\\
&&+2\left(\frac{\lambda_{x}g_{x}}{2}\right)^{2}\left[\frac{D-\Omega}{\left(D-\Omega\right)^{2}-\xi_{z}^{2}}+a^{\dagger}a\left(\frac{D-\xi_{z}}{\left(D-\xi_{z}\right)^{2}-\Omega^{2}}+\frac{D+\xi_{z}}{\left(D+\xi_{z}\right)^{2}-\Omega^{2}}\right)\right]\left(X^{-,+}+X^{+,-}\right)
\end{eqnarray}
where as described in the main text, we have neglected second order photon terms, but we have now kept the second order transition operators $X^{\pm,\mp}$ to check the deviation from a QND measurement. Importantly, one can see how each spin subspace is affected differently by the coupling with the cavity, due to the influence of the non-linear anisotropy $D$. From this expression it is easy to extract the frequency shifts described in the main text.

Finally, we can calculate the deviation from a perfect QND measurement by calculating the commutation between the unperturbed spin Hamiltonian and the frequency shift term. It yields:
\begin{eqnarray}
\sum_{\beta_{i}=1}^{2S+1}E_{\vec{\beta}}\Phi_{\vec{\beta}}X^{\vec{\beta}}&=&2\left(\frac{\lambda_{x}g_{x}}{2}\right)^{2}\left(\frac{E_{+}}{E_{+}^{2}-\Omega^{2}}+\frac{E_{-}}{E_{-}^{2}-\Omega^{2}}\right)\left(E_{+,-}X^{+,-}+E_{-,+}X^{-,+}\right)\nonumber\\
&=&4\xi_{z}\left(\frac{\lambda_{x}g_{x}}{2}\right)^{2}\left(\frac{D+\xi_{z}}{\left(D+\xi_{z}\right)^{2}-\Omega^{2}}+\frac{D-\xi_{z}}{\left(D-\xi_{z}\right)^{2}-\Omega^{2}}\right)\left(X^{+,-}-X^{-,+}\right)
\end{eqnarray}
where we have used:
\begin{equation}
    \Phi_{\vec{\beta}}=\sum_{\alpha=1}^{2S+1}\Lambda_{\beta_{1},\alpha}\Lambda_{\alpha,\beta_{2}}\left(\frac{E_{\beta_{1},\alpha}}{E_{\beta_{1},\alpha}^{2}-\Omega^{2}}+\frac{E_{\beta_{2},\alpha}}{E_{\beta_{2},\alpha}^{2}-\Omega^{2}}\right)
\end{equation}
Notice that if $D\to 0$, the commutator vanishes, indicating that the measurement is QND, to second order in $\Lambda_{\vec{\alpha}}$. Therefore, we can conclude that even for this case with purely longitudinal anisotropy and fully transverse interaction, the measurement is not QND due to the presence of the non-linear term $(S^z)^2$. Nevertheless, the non-commutativity is proportional to $(\lambda_x g_x)^2$, which is small, times the second order transition operators $X^{\pm,\mp}$, whose expectation value is also small in the weak coupling regime. Hence, for all practical purposes, it might be possible to neglect this effect during the readout, but nevertheless it should be estimated.

\section{Heterodimetallic {[}CeEr{]} Lanthanide Complex}
\label{app:ceer}

This molecule has the special feature of being an ionic dimer. From
the experiment it is possible to find the parameters for the diagonal form of
the $\hat{g}_{1,2}$ tensor in each ion independently  by fitting. However, the
molecule accommodates the two ions with a relative orientation which
has been estimated to be of the order of $\theta=70\text{º}$. This
implies that the calculation of the eigenstates of the isolated molecule
requires to rotate $\hat{g}_{2}$ for the Ce ion. We consider a rotation
in the x-z plane implemented by the matrix:
\begin{equation}
\hat{R}=\left(\begin{array}{ccc}
\cos\theta & 0 & \sin\theta\\
0 & 1 & 0\\
-\sin\theta & 0 & \cos\theta
\end{array}\right)
\end{equation}
which transforms $\hat{g}_{2}$ into the following form:
\begin{equation}
\hat{g}_{2}=\left(\begin{array}{ccc}
g_{2}^{x}\cos^{2}\theta+g_{2}^{z}\sin^{2}\theta & 0 & \left(g_{2}^{z}-g_{2}^{x}\right)\cos\theta\sin\theta\\
0 & g_{2}^{y} & 0\\
\left(g_{2}^{z}-g_{2}^{x}\right)\cos\theta\sin\theta & 0 & g_{2}^{z}\cos^{2}\theta+g_{2}^{x}\sin^{2}\theta
\end{array}\right)
\end{equation}
This allows us to write the Zeeman term as:
\begin{align}
H_{Z}= & -\mu_{B}B_{z}\left[g_{1}^{z}S_{1}^{z}+\left(g_{2}^{z}\cos^{2}\theta+g_{2}^{x}\sin^{2}\theta\right)S_{2}^{z}\right]\nonumber \\
 & -\mu_{B}B_{z}\left(g_{2}^{z}-g_{2}^{x}\right)\cos\theta\sin\theta S_{2}^{x}
\end{align}
In addition, we assume that the interaction tensor is scalar (lowest
order approximation, where each spin reacts to the effective magnetic
field produced by the other), which means that the interaction Hamiltonian
can be written as: 
\begin{eqnarray}
V & = & -\frac{J_{12}}{g_{J1}g_{J2}}g_{1}^{x}\left(g_{2}^{x}\cos^{2}\theta+g_{2}^{z}\sin^{2}\theta\right)S_{1}^{x}S_{2}^{x}\\
 &  & -\frac{J_{12}}{g_{J1}g_{J2}}g_{1}^{y}g_{2}^{y}S_{1}^{y}S_{2}^{y}\nonumber \\
 &  & -\frac{J_{12}}{g_{J1}g_{J2}}g_{1}^{z}\left(g_{2}^{z}\cos^{2}\theta+g_{2}^{x}\sin^{2}\theta\right)S_{1}^{z}S_{2}^{z}\nonumber \\
 &  & -\frac{J_{12}}{g_{J1}g_{J2}}\cos\theta\sin\theta\left(g_{2}^{z}-g_{2}^{x}\right)\left(g_{1}^{x}S_{1}^{x}S_{2}^{z}+g_{1}^{z}S_{1}^{z}S_{2}^{x}\right)\nonumber 
\end{eqnarray}
Finally, the coupling with the cavity photons must include the relative
angle between the two ions, resulting in the following form:
\begin{eqnarray}
\sum_{i}\vec{\epsilon}\cdot\hat{g}_{i}\cdot\vec{S}_{i} & = & \epsilon_{x}g_{1}^{x}S_{1}^{x}+\epsilon_{y}\left(g_{1}^{y}S_{1}^{y}+g_{2}^{y}S_{2}^{y}\right)+\epsilon_{z}g_{1}^{z}S_{1}^{z}\\
 &  & +\left(g_{2}^{x}\cos^{2}\theta+g_{2}^{z}\sin^{2}\theta\right)\epsilon_{x}S_{2}^{x}\nonumber \\
 &  & +\left(g_{2}^{z}\cos^{2}\theta+g_{2}^{x}\sin^{2}\theta\right)\epsilon_{z}S_{2}^{z}\nonumber \\
 &  & +\left(g_{2}^{z}-g_{2}^{x}\right)\cos\theta\sin\theta\left(\epsilon_{z}S_{2}^{x}+\epsilon_{x}S_{2}^{z}\right)\nonumber 
\end{eqnarray}
These are the main manipulations used to obtain the effective Hamiltonian for the dimer coupled to the cavity photons.
\end{widetext}
\end{document}